%% file: main.tex
\documentclass[sigconf, 9pt]{acmart}

\pdfoutput=1

\input preamble/preamble
\input preamble/macros

\author{Myrsini Gkolemi}
\affiliation{
	\institution{FORTH \& University of Crete}
	\country{Greece}
}
\author{Panagiotis Papadopoulos}
\affiliation{
	\institution{Telefonica Research}
	\country{Spain}
}
\author{Evangelos P. Markatos}
\affiliation{
	\institution{FORTH \& University of Crete}
	\country{Greece}
}
\author{Nicolas Kourtellis}
\affiliation{
	\institution{Telefonica Research}
	\country{Spain}
}
\begin{document}

\title{YouTubers Not \emph{madeForKids}: Detecting Channels Sharing Inappropriate Videos Targeting Children}

\copyrightyear{2022}
\acmYear{2022}
\setcopyright{acmcopyright}
\acmConference[WebSci '22]{14th ACM Web Science Conference 2022}{June 26--29, 2022}{Barcelona, Spain}
%\acmBooktitle{14th ACM Web Science Conference 2022 (WebSci '22), June 26--29, 2022, Barcelona, Spain}
\acmPrice{15.00}
\acmDOI{10.1145/3501247.3531556}
\acmISBN{978-1-4503-9191-7/22/06}

\input sections/00_abstract
\maketitle

\input sections/01_introduction

\input sections/03_dataset

\input sections/09_example
\input sections/04_measurements

\input sections/05_classifier
\input sections/07_related

\input sections/08_conclusion

\begin{acks}
 This project received funding from the EU H2020 Research and Innovation programme under grant agreements No 830927 (Concordia), No 830929 (CyberSec4Europe), No 871370 (Pimcity) and No 871793 (Accordion).
These  results reflect only the authors' view and the Commission is not responsible for any use that may be made of the information it contains.
\end{acks}

\bibliographystyle{ACM-Reference-Format}
\balance
\bibliography{main}

\appendix
\input sections/appendix

\end{document}

%% file: preamble/preamble.tex
\usepackage{booktabs}  % for toprule, bottomrule in tables
\usepackage{endnotes}
\usepackage{epstopdf}
\usepackage{graphicx}
\usepackage{xspace}
\usepackage{enumitem}
\usepackage{url}
\usepackage{balance}
\usepackage[utf8]{inputenc}
\usepackage{array,tabularx}
\usepackage{multirow}
\usepackage{hyperref}

\usepackage[tableposition=top,font={footnotesize, bf},figurename=Figure, skip=7pt]{caption}     % captions on top for tables

%% file: preamble/macros.tex
\newcommand{\etc}{etc.\xspace}

\newcommand{\ie}{i.e.,\xspace}

\newcommand{\point}[1]{\vspace{.05in} \par\noindent\textbf{#1}:\xspace}

\newcommand{\emojiregistered}{\includegraphics[width=1.8em]{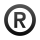}}
\newcommand{\emojiredheart}{\includegraphics[width=1.8em]{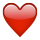}}
\newcommand{\emojitrademarksign}{\includegraphics[width=1.8em]{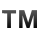}}
\newcommand{\emojiheartsuit}{\includegraphics[width=1.8em]{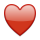}}
\newcommand{\emojibabychick}{\includegraphics[width=1.8em]{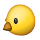}}
\newcommand{\emojismilingfaceheartshapedeyes}{\includegraphics[width=1.8em]{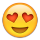}}
\newcommand{\emojifacewithtearsofjoy}{\includegraphics[width=1.8em]{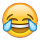}}
\newcommand{\emojibiohazard}{\includegraphics[width=1.8em]{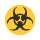}}
\newcommand{\emojimalesign}{\includegraphics[width=1.8em]{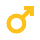}}
\newcommand{\emojibackhandindexpointingdown}{\includegraphics[width=1.8em]{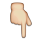}}
\newcommand{\emojigreatbritain}{\includegraphics[width=1.8em]{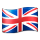}}
\newcommand{\emojicopyright}{\includegraphics[width=1.8em]{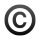}}
\newcommand{\emojiexclamationmark}{\includegraphics[width=1.8em]{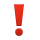}}

%% file: sections/00_abstract.tex
\begin{abstract}
YouTube is one of the most popular social media and online video sharing platforms, and users turn to it for entertainment by consuming music videos, for educational or political purposes, advertising, etc.
In the last years, hundreds of new channels have been creating and sharing videos targeting children, with themes related to animation, superhero movies, comics, etc.
Unfortunately, many of these videos have been found to be inappropriate for consumption by their target audience, due to disturbing, violent, or sexual scenes.

In this paper, we study YouTube channels that were found to post \emph{suitable} or \emph{disturbing} videos targeting kids in the past.
Unfortunately, we identify a clear discrepancy between what YouTube assumes and flags as inappropriate content and channel, vs. what is found to be disturbing content and still available on the platform, targeting kids.
In particular, we find that almost 60\% of videos that were manually annotated and classified as \emph{disturbing} by an older study in 2019 (a collection bootstrapped with \emph{Elsa} and other keywords related to children videos), are still available on YouTube in mid 2021.
In the meantime, 44\% of channels that uploaded such \emph{disturbing} videos, have yet to be suspended and their videos to be removed. 
For the first time in literature, we also study the ``madeForKids'' flag, a new feature that YouTube introduced in the end of 2019, and compare its application to the channels that shared
\emph{disturbing} videos, as flagged from the previous study.
Apparently, these channels are less likely to be set as ``madeForKids'' than those sharing suitable content.
In addition, channels posting \emph{disturbing} videos utilize their channel features such as keywords, description, topics, posts, etc., in a way that they appeal to kids (e.g., using game-related keywords).
Finally, we use a collection of such channel and content features to train machine learning classifiers that are able to detect, at channel creation time, when a channel will be related to \emph{disturbing} content uploads. 
These classifiers can help YouTube content moderators reduce such incidences, by pointing to potentially suspicious accounts, without analyzing actual videos, but instead only using channel characteristics.
\end{abstract}

%% file: sections/01_introduction.tex
\section{Introduction}
\label{sec:intro}

In the last few years, there has been a dramatic increase in the use of social media, and especially platforms for video sharing and consumption such as TikTok and YouTube~\cite{growth-social-media,iqbal2020youtubestats}.
In fact, this has been the case during COVID-19 lockdowns~\cite{covid-19-lockdowns}, with a general increase in daily and monthly active users~\cite{chaffey2020worldstats-social-media,fischer2020worldstats-social-media-2}, and YouTube and Facebook-based content being among the most shared~\cite{bottger2020covid19-internetreaction-imc, kemp2020worldstats-social-media3}.

Nonetheless, along with the generation and exposure to valuable and acceptable content, 
there have been frequent uploads of media which are deemed inappropriate for specific audiences.
This is an important issue regarding YouTube videos, which in spite of presenting kid-related topics (e.g., cartoons, animation movies, etc.), they can often be inappropriate for children, as the videos display disturbing, violent, sexual or other improper scenes~\cite{elsagate,papadamou2020disturbed-youtube-for-kids}.
This problem has been of particular importance during recent COVID-related restrictions and confinements, since many parents resort to video platforms, such as YouTube and TV programs, to keep their children occupied while schools are closed.
Consequently, children end up spending many hours per day watching videos, some of which could be inappropriate~\cite{children-social-media, ofcom-findings}.

In order to address this ongoing problem, YouTube has proceeded to apply various methods and filtering in the last few years.
Among them are: (i)~a system of 3 strikes that forces the channel owner to be careful what they upload or make available on their channel, as they could be banned from the platform~\cite{guidelines-strike}, (ii)~a \emph{Trusted Flaggers} program~\cite{trusted-flaggers} in which individual users, government agencies and NGOs notify YouTube of content that violates the Community Guidelines, (iii)~machine learning methods for detecting inappropriate content~\cite{youtube-machine-learning}, (iv)~a specialized YouTube platform making available content only for kids~\cite{youtube-kids}, and (v)~a recently introduced flag, ``madeForKids'' ~\cite{madeforkids-coppa}, that allows creators to declare whether their content is kid-appropriate or not. This is not only useful for better promoting and recommending content to users searching for kid-related videos, but also accelerates auditing of such videos by YouTube algorithms and moderators~\cite{made-for-kids-flag}.

Past research has examined the problem from a video content point of view, and analyzed features available on videos and channels such as comments posted, number of views, thumbnails, and even video snapshots~\cite{papadamou2020disturbed-youtube-for-kids,ishikawa2019elsagate-phenomenon,tahir2019kid-youtube-kids,han2020elsagate}.
However, they have not addressed the problem from the perspective of accounts who post such videos, and whether their various characteristics reveal a tendency for posting \emph{suitable} or \emph{disturbing} videos.

In this paper, we make the following contributions:
\begin{itemize}[nolistsep]
    \item We are the first to study the characteristics of YouTube accounts that publish inappropriate videos targeting kids. In particular, we look into how older videos and accounts have been banned by YouTube for violating its policies on content publishing. We find that only 28.5\% of channels that have uploaded disturbing content (and have been assessed as such in 2019) have been terminated by YouTube by mid 2021. In fact, almost 60\% (or 546) of manually annotated disturbing videos are still accessible through the platform by mid 2021.
    \item We study the newly added flag from YouTube called ``madeForKids'' to understand its association to the inappropriate content and accounts publishing it. We discover that 25\% of channels with suitable content are set to ``madeForKids'', while only 3\% of channels with inappropriate content are set as such.
    \item We analyze 27 different characteristics of channels and how these features are associated with the type of channel and the content it publishes (\ie if it was found to be disturbing or suitable for kids).
    Among these features are country and channel creation date, statistics like subscriptions and video views, keywords and topics, social media links, polarity and sentiment of description etc.
    \item Finally, we demonstrate how these features can help YouTube build a basic machine learning classifier to infer if a channel is likely to share inappropriate/disturbing videos or not, with up to $AUC=0.873$. In fact, we show how this is possible to perform even at channel creation time, by using only features available at that moment and disregarding historical activity features, with up to $AUC=0.869$.
    \item We make our data and code available for research reproducibility and extensibility.\footnote{\url{{https://github.com/Mirtia/Inappropriate-YouTube}}}
\end{itemize}

%% file: sections/03_dataset.tex
\section{Data Collection}
\subsection{YouTube Crawling \& Feature Extraction}
\label{sec:data-crawling}

\begin{figure*}[t]
    \centering
    \includegraphics[width=2.1\columnwidth]{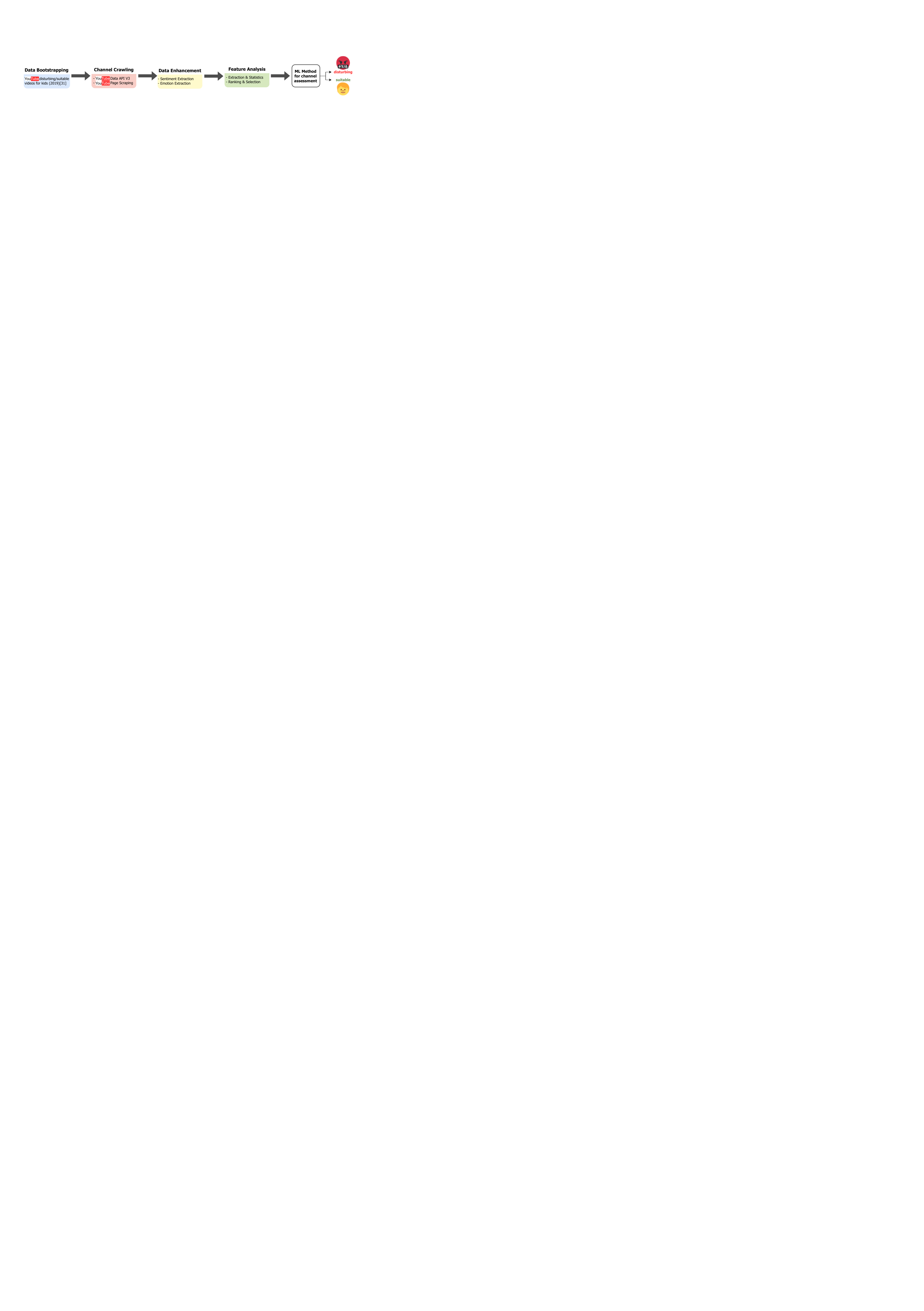}
    \caption{Overview of methodology followed in this study.}
    \label{fig:methodology}
\end{figure*}

The methodology followed in this study is outlined in Figure~\ref{fig:methodology}.
We begin by studying the data made available from a past study~\cite{papadamou2020disturbed-youtube-for-kids} on the topic. 
The past ground truth dataset which was randomly sampled by a set of 844K videos assembled by /r/Elsagate and /r/fullcartoonsonyoutube~\cite{fullcartoons-subreddit} subreddits, includes details of 4797 YouTube videos and their four associated labels as provided by human annotators: \emph{disturbing, suitable, restricted} and \emph{irrelevant}. Each video was annotated by two of the authors of~\cite{papadamou2020disturbed-youtube-for-kids} and one undergraduate student with the assistance of a platform that includes a clear description
of the annotation task, the labels, as well as all the video information needed for the inspection.
Since our focus is videos that target children, we ignore the videos with labels \emph{restricted} and \emph{irrelevant}, and analyze the channels that posted 2442 videos with labels \emph{suitable} or \emph{disturbing}.
We call this subset the $GT$ dataset.
Features are divided into three categories according to the crawling method or channel section they belong to. In Table~\ref{tab:features-tab}, it is clear that most features were collected via YouTube API v3.

\point{YouTube Data API v3}
First step in our data crawling process was to revisit these videos with YouTube's Data API v3, and assess their status (\ie~if they are available or not), as well as collect further public information about channels that published these videos.
Each channel is distinguished by a unique 24-character identifier.
To reach a channel, you ``concat'' the identifier with the specified (URLs): \emph{https://www.youtube.com/channel/ID,  https://www.youtube.com/c/ID}.

In particular, during this crawling, we collected the status and following attributes associated with each channel: ``country'', ``description'', ``keywords'', ``publishedAt``, ``madeForKids'', ``topicCategories'', `viewCount'', ``videoCount'', ``subscriberCount'', as well as calculated counts such as ``keywordsCount'', ``topicCount'', ``subscriptionCount'', ``descriptionCharCount'' and ``postCount''. 
For the sake of clarification, ``publishedAt'' states the date a YouTube channel joined the platform and ``topicCategories'' is a list of Wikipedia URLs that describe the channel's content.
We note that since YouTube Data API v3 did not provide a method to parse the status of each video or channel, we used the \emph{Beautiful Soup Python Library}~\cite{beautiful-soup} instead, to scrape the relative messages from the page source.
Ethical considerations of our crawling method are addressed in Appendix~\ref{sec:ethics}.

\begin{table}[t]
\caption{Data collected from YouTube channels.}
\centering
\footnotesize
    \begin{tabular}{ll}
        \toprule
        \textbf{Source} &\textbf{Features Collected}\\ 
        \midrule
        YouTube         & country, description, keywords, topicCategories, datePublished,\\
        API             & madeForKids, viewCount, videoCount, subscriberCount,\\
        derived         & postCount, subscriptionCount, hiddenSubsribersCount(boolean),\\
                        & linksCount, descriptionCharCount, topicCount, subscriptionsList\\
        \hline
        Community       & datePublished, description, tags, hashtags, externalLinks,\\
        Tab Post        & youtubeLinks, channelLinks, likeCount, thumbnailVideo\\
        \hline
        About Tab & email, links (text, URL)\\
        \bottomrule
    \end{tabular}
    \label{tab:features-tab}\vspace{-0.3cm}
\end{table}  

\point{Community and About Tabs}
Apart from these features, we also inspected other publicly available sources of account-centered information, such as the ``Community Tab'' and ``About Tab''.
The Community Tab contains posts with enriched media uploaded by the account owner.
As this is a newly added feature, YouTube Data API v3 does not offer a method to get its information automatically.
Therefore, in order to collect these posts, we used \emph{Puppeteer}~\cite{puppeteer} and Python's \emph{concurrent.futures}~\cite{python-concurrent} for multi-threading, along with Beautiful Soup to scrape the resulting pages at a limited request rate that may not disturb the YouTube platform.
We focused on 100 posts of each channel as an indicator of what type of content the channel owner generally posts.
Features extracted per post are: ``datePublished'', ``description'', ``tags'', ``hashtags'', ``externalLinks'', ``youtubeLinks'', ``channelLinks'', ``likeCount'', and ``thumbnailVideo''.
In particular, ``channelLinks'' are URLs of other tagged channels or users in the description; ``externalLinks'' are URLs found in the description and redirect to other pages than YouTube; ``thumbnailVideo'' is the ID of the video embedded in a post.
The About Tab of a channel consists of a description section, details (email for business inquiries, location), stats (date the user joined YouTube, number of views) and links (social media, merchandise, \etc).
We used Puppeteer to collect both links and emails.

\point{Sentiment \& Emotion Extraction:}
In order to extract features related to \emph{sentiment} and \emph{emotion}, we used the \emph{MeaningCloud Deep Categorization API Emotion Detection}~\cite{meaningcloud} to classify the text description of each channel.
In addition to Emotion detection, we calculated polarity of keywords, posts and channel description using the well-known \emph{SentiStrength}~\cite{sentistrength} library.

\subsection{Channel Labeling}
\label{sec:data-labels}

As mentioned earlier, the videos were split into four categories: \emph{disturbing, suitable, restricted} and \emph{irrelevant}.
We focus on \emph{suitable} and \emph{disturbing}, depending on whether the content shown is appropriate or not for children.

These two labels were introduced in the past study on the subject of detecting disturbing YouTube videos for kids.
Any video that is not age-restricted but targets children audience and contains sexual hints, horror scenes, inappropriate language, graphic nudity and child abuse was labeled as \emph{disturbing}.
According to YouTube Child safety policy~\cite{child-safety-policy}, a video would be considered inappropriate(\emph{disturbing}) if it contains misleading family content, cyber-bullying and harassment involving minors.
On the other hand, a video is \emph{suitable} when its content is appropriate for children (G-rated~\cite{motion-picture-association-rating}) and it is relevant to their typical interests.
We consider a channel ``potentially disturbing’’ when they have already uploaded at least one video that was manually annotated as \emph{disturbing} by the previous study.
For sake of simplicity, we refer to these channels as \emph{disturbing} for the rest of the study.

\begin{figure}[t]
    \centering
    \includegraphics[width=0.67\columnwidth]{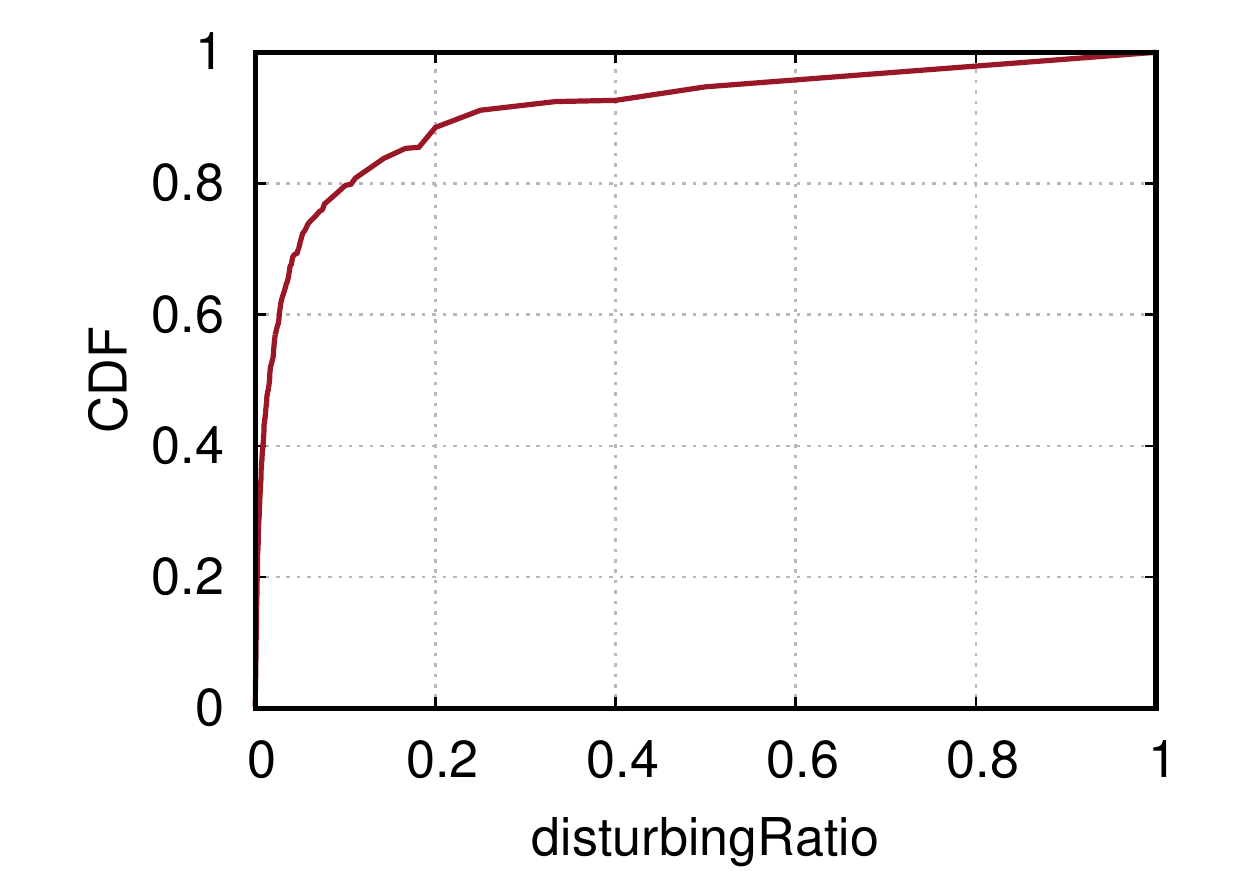}
    \caption{CDF of disturbingRatio, \ie~number of disturbing videos found in an channel over the total number of videos (suitable+disturbing) from that channel, when that channel had at least 1 disturbing video.}
    \label{fig:disturbingRatio}
\end{figure}

Then, we look into the number of \emph{disturbing} videos that each channel posted, from $GT$.
Figure~\ref{fig:disturbingRatio} plots the CDF of the ratio of disturbing videos to total videos within $GT$, per channel that had at least one disturbing video in the original dataset.
Through YouTube v3 API, we confirm that $\sim$5\% of accounts with reported disturbing videos have zero ``videoCount'' because they were probably unlisted, privatized or reported for violation of YouTube Guidelines.

Based on this preliminary result, we make the following assumptions when propagating the video labels to the channels:
\begin{itemize}
    \item \textbf{Suitable Channel:} If it has published only ``suitable'' videos, based on the videos in $GT$.
    \item \textbf{Disturbing Channel:} If it has published at least one ``disturbing'' video, based on the videos in $GT$.
\end{itemize}
Table~\ref{tab:dataset} summarizes the number of videos and channels from our crawls, along with their associated labels which we use in the rest of the study.
All crawls on YouTube were performed in mid 2021.

\begin{table}[t]
    \centering
    \caption{Number of videos and channels per label.
    \emph{Total} reflects the number of videos (and consequently channels) that were originally in the $GT$ dataset.
    \emph{Available} reflects the videos and channels that were successfully crawled in 2021 and are studied in this paper.}
    \scalebox{0.8}{
    \begin{tabular}{lrrrr}
    \toprule
    \multirow{2}{*}{\textbf{Category}} &
    \multicolumn{2}{c}{\textbf{Channels}} &
    \multicolumn{2}{c}{\textbf{Videos}} \\
    & \bf Total & \bf Available & \bf Total & \bf Available \\
    \midrule
    suitable             &   909    &   779 &   1513    &   1505 \\ 
    disturbing           &   789    &   559 &   929     &   539  \\ 
    \bottomrule
    \end{tabular}
    }\label{tab:dataset}\vspace{-0.3cm}
\end{table}

\begin{figure*}[t]
    \centering
    \includegraphics[width=\textwidth]{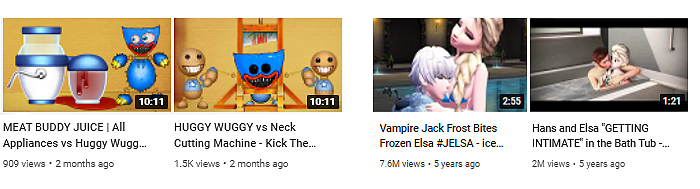}
    \caption{Overview of methodology followed in this study.}
    \label{fig:example-disturbing}
\end{figure*}

%% file: sections/09_example.tex
\subsection{Examples of Disturbing Channels}

Inappropriate content comes into various forms, from a grotesque clickbait thumbnail to horror stories with cartoon characters. 
For the sake of example, we provide thumbnails of videos that some channels we labelled as ``disturbing'' have been hosting in their accounts.
Please note that these videos were still available on May 2022, i.e., more than one year after the initial YouTube crawls of our aforementioned dataset, and two years after the initial dataset of inappropriate videos for kids was published~\cite{papadamou2020disturbed-youtube-for-kids}.

Figure~\ref{fig:example-disturbing} shows various examples (via screenshots) of such inappropriate content targeting kids.
To the left side of Figure~\ref{fig:example-disturbing}, there is an example of a channel uploading gameplay videos to promote games for children.
The thumbnails depict a doll getting tortured with various tools.
On the right side of Figure~\ref{fig:example-disturbing}, we can see another channel included in the dataset, which uploads implied sexual content of animated characters, mainly Elsa.
Other examples, omitted here due to space, include horror parodies of Peppa the Pig and videos with actors role-playing as famous comic characters that engage into explicit acts. 

%% file: sections/04_measurements.tex
\section{Channel Feature Analysis}

\subsection{Why are videos and channels removed?}
\label{sec:removal-reason}

First, we look into the status of videos annotated by the past study, as well as the accounts that posted them.
This is important in order to assess which videos from the \emph{disturbing} set may have been removed by YouTube, and in what extent the reasoning behind the removal aligns with the label provided by the past study.
Whenever a video is not available in the platform, YouTube displays a characteristic message explaining the reason why the user cannot view the video. Since YouTube API v3 does not include methods to collect error messages on removed videos, we used Beautiful Soup to parse them.
In general, YouTube videos may not be reachable because of different reasons: unavailability of the service or network (less likely), if the content was made private by the owner, or if the video was going against the Community guidelines and policies of YouTube and was removed.

\begin{figure}[t]
\centering
    \begin{minipage}[t]{\columnwidth}
    \centering
    \includegraphics[width=\textwidth]{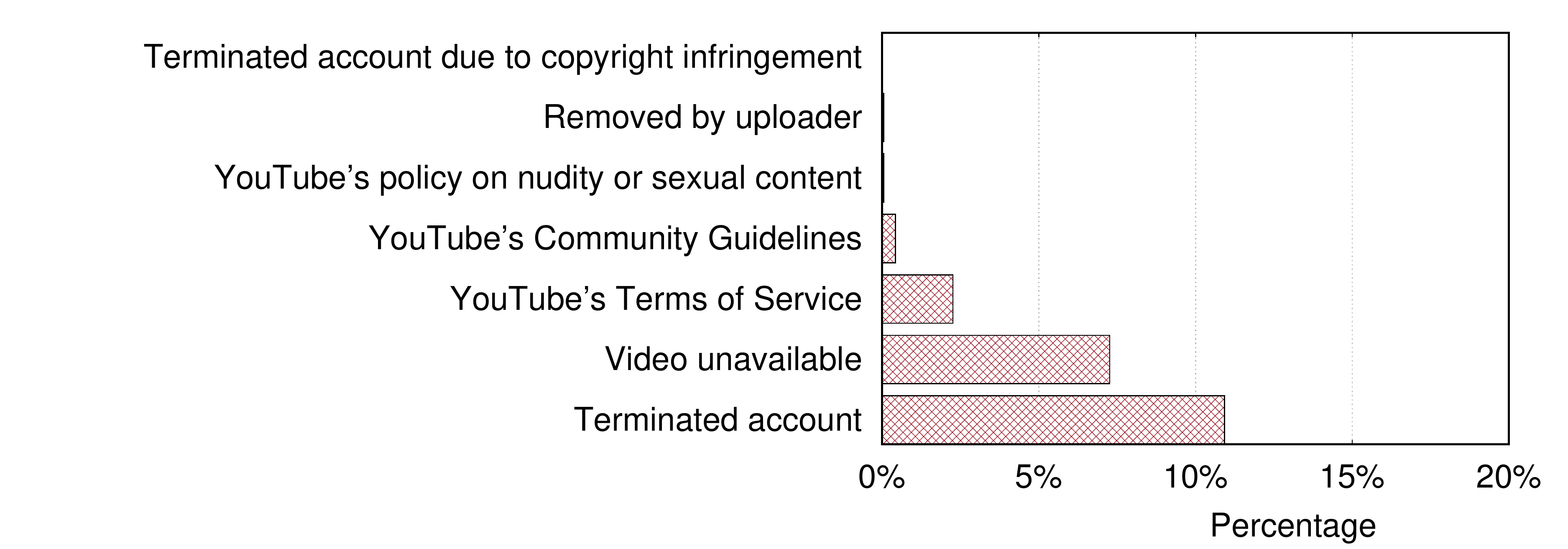}
    \caption{Reasons why YouTube videos labeled as ``disturbing'' are not currently available on the platform (\% total videos in $GT$).}
    \label{fig:reasonsDVideos}
\end{minipage}
\end{figure}

We analyze the reasons why videos classified as ``disturbing'' or ``suitable'' were removed by YouTube.
As shown in Table~\ref{tab:dataset}, only 0.1\% of \emph{suitable} videos were removed, while more than 40\% of \emph{disturbing} videos were taken down, with the dominant reason being account termination.
More specifically, and as shown in Figure~\ref{fig:reasonsDVideos}, 10.9\% (203) of removed disturbing videos are linked with terminated accounts and 2.2\% of such videos are linked with accounts banned because of not respecting YouTube Terms of Service.

\begin{figure}[t]
    \begin{minipage}[t]{\columnwidth}
    \centering
    \includegraphics[width=\textwidth]{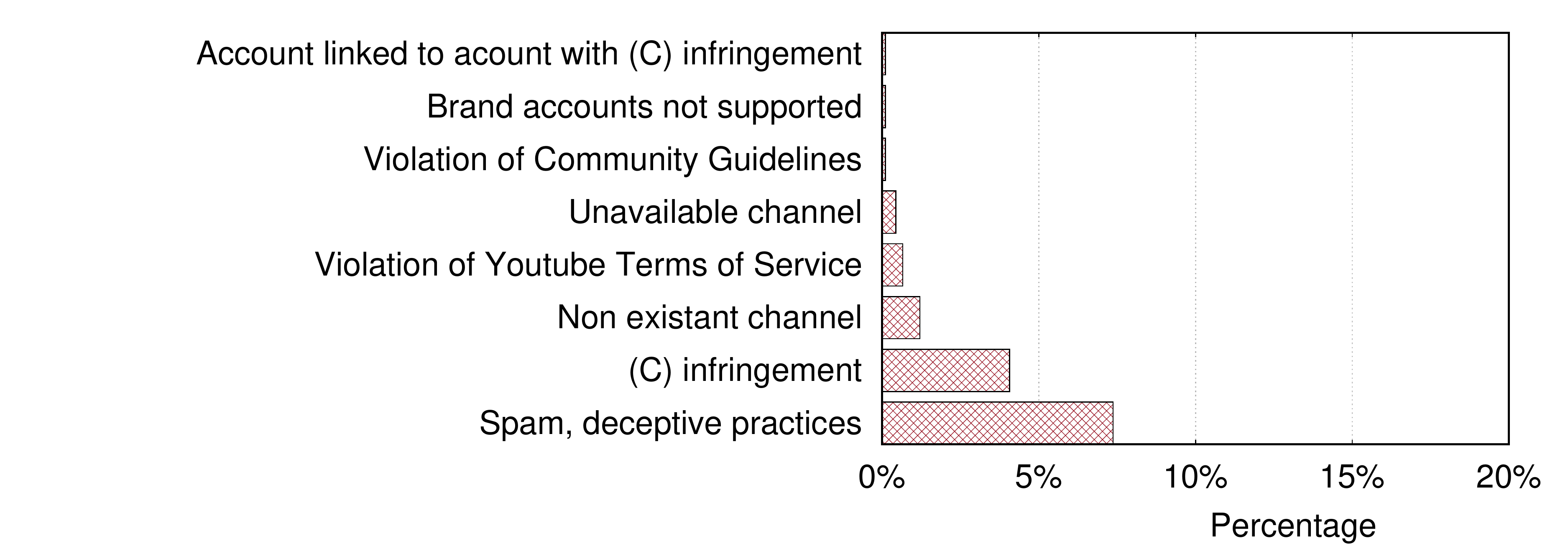}
    \caption{Reasons why YouTube channels labeled as ``suitable'' are not currently reachable on the platform (\% total channels in $GT$).}
    \label{fig:reasonsSChannels}
\end{minipage}
\end{figure}
\begin{figure}[t]
    \begin{minipage}[t]{\columnwidth}
    \centering
    \includegraphics[width=\textwidth]{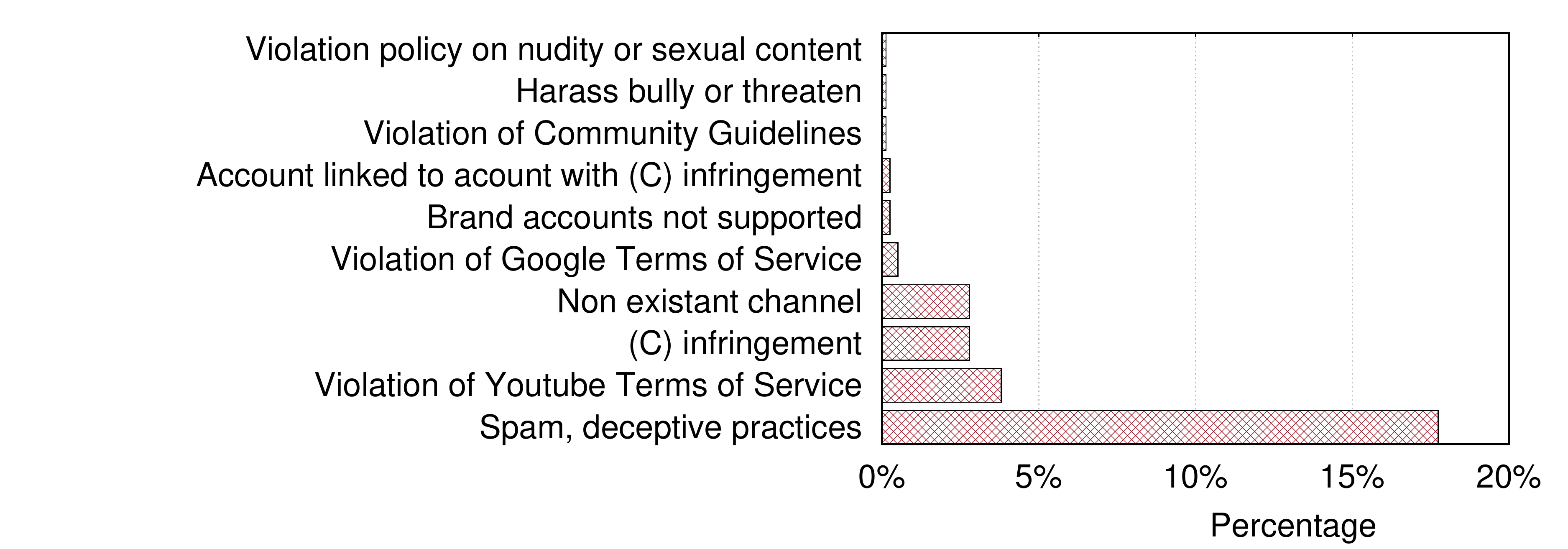}
    \caption{Reasons why YouTube channels labeled as ``disturbing'' are not currently reachable on the platform (\% total channels in $GT$).}
    \label{fig:reasonsDChannels}
    \end{minipage}
\end{figure}

After studying the possible causes of why videos were taken down, we move to examine the status of channels that uploaded these videos.
This data collection consists of each channel and their respective videos included in $GT$.
YouTube actions on violating Community Guidelines consist of four levels~\cite{guidelines-strike}.
In the beginning, the user who owns the account receives a warning, apart from severe abuse cases when the channel is terminated immediately.
The second time a user's content is considered improper, they receive a strike.
Their actions, such as uploading videos, creating or editing playlists, etc., are restricted for a week.
However, the strike remains on the channel for 90 days.
In case the user receives a second strike during this period, they become incapable of posting content for two weeks.
A third strike during this time interval results in permanent removal of the channel.

As we see in Figure~\ref{fig:reasonsSChannels}, suitable channels were less likely to have been removed during the elapsed time between the past study in our crawls.
In fact, 7.37\% of suitable channels were terminated due to multiple small or severe violations of YouTube's policy against spam, deceptive practices, and misleading content, or other Terms of Service violations, and 4.07\% in consequence of copyright infringement.
Instead, in Figure~\ref{fig:reasonsDChannels}, we observe that more than double (17.74\%) of disturbing channels were banned from YouTube platform because of spam and deceptive practice policies, as well as for violating YouTube Terms of Service (3.8\%), copyright infringement (2.78\%) channel absence (2.78\%). 

Overall, and after our crawls and analysis, while 929 videos were classified in the past study as ``disturbing'', 58.8\% are still reachable in mid 2021.
In fact, only 28.5\% of the users/channels that have uploaded such disturbing content have been terminated by YouTube, demonstrating a lack of action by the platform.

\subsection{Are videos and channels \emph{MadeForKids}?}
\label{sec:made-for-kids}

YouTube Creators published a video on the updates of ``Complying with COPPA'' on 12th of November, 2019~\cite{madeforkids-coppa} where they introduced the ``madeForKids'' label for both channels and videos.
This feature denotes whether the content of a video or channel is directed at children.
More specifically, the content is ``madeForKids'' if it is child-friendly, and most likely includes child actors, animated characters or cartoon figures, or serves educational purposes.

To comply with the Children’s Online Privacy Protection Act (COPPA)~\cite{coppa-wiki} and other related laws, YouTube makes certain features of its regular channels unavailable on ``made for Kids'' content and channels.
Regarding videos, these switched-off features include: auto-play on home, cards or end screens, channel branding watermark, comments, donate button, likes/dislikes on YouTube Music, live chat or live chat donations, merchandise and ticketing, notifications, personalized advertising, playback in the Mini-player, Super Chat or Super Stickers, save to playlist and save to watch later.
At channel level, the restricted features include Channel Memberships, Notifications, Posts, and Stories. 
Regarding the aforementioned ``madeForKids'' flag, a channel can be:
\begin{enumerate}[nolistsep]
    \item ``madeForKids'': allowed to only post videos ``madeForKids'';
    \item not ``madeForKids'': allowed to only post videos that are not ``madeForKids'';
    \item not defined: each video is defined if it is ``madeForKids'' or not on upload time;
\end{enumerate}
However, YouTube is also supported by a machine learning algorithm to detect incorrectly labeled videos and set them according to their content~\cite{madeforkids-coppa}.

\begin{figure}[t]
\begin{minipage}[t]{0.23\textwidth}
    \centering
    \includegraphics[width=1.0\linewidth]{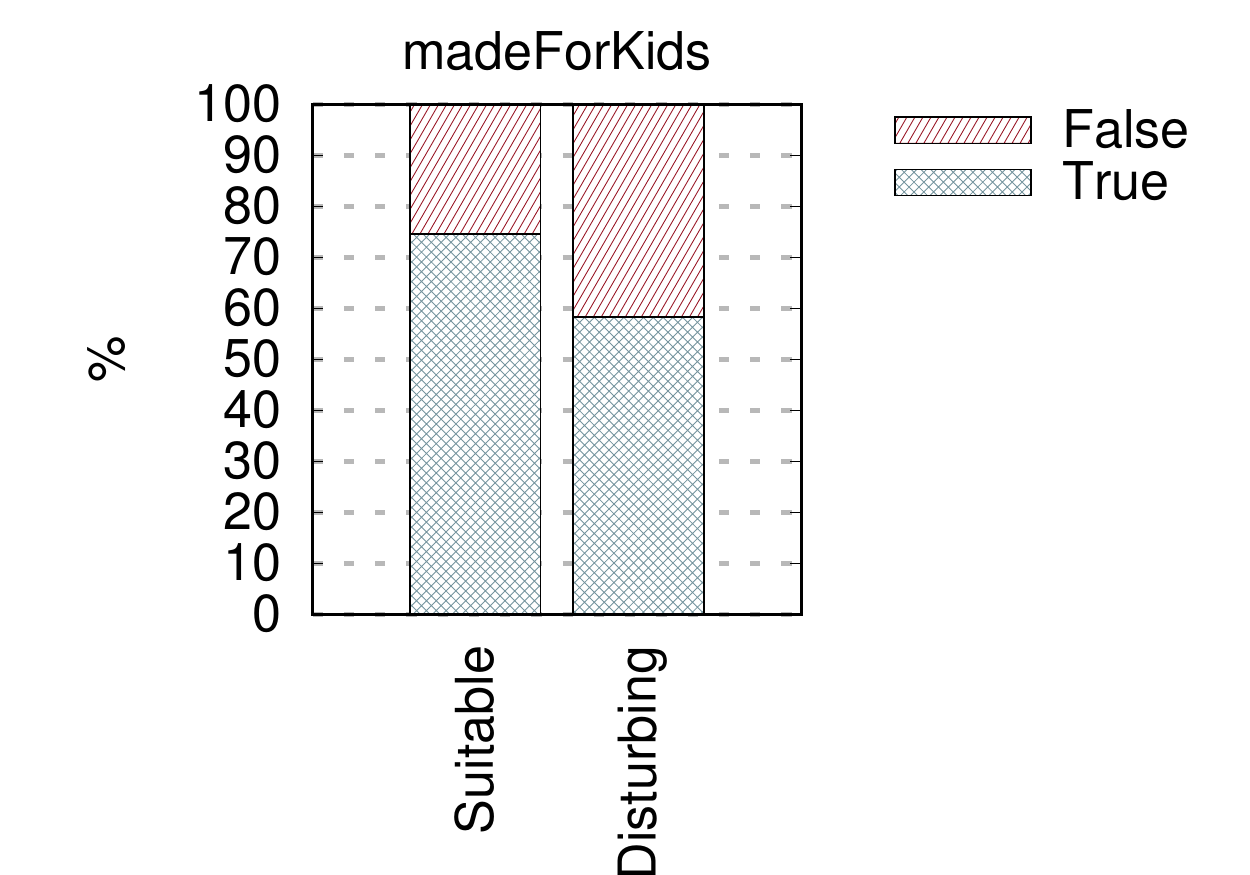}
    \caption{The use of \emph{madeForKids} label by videos on YouTube labeled as suitable or disturbing.}
    \label{fig:madeForKidsVideos}
    \end{minipage}
    \hfill
    \begin{minipage}[t]{0.23\textwidth}
    \centering
    \includegraphics[width=1.0\linewidth]{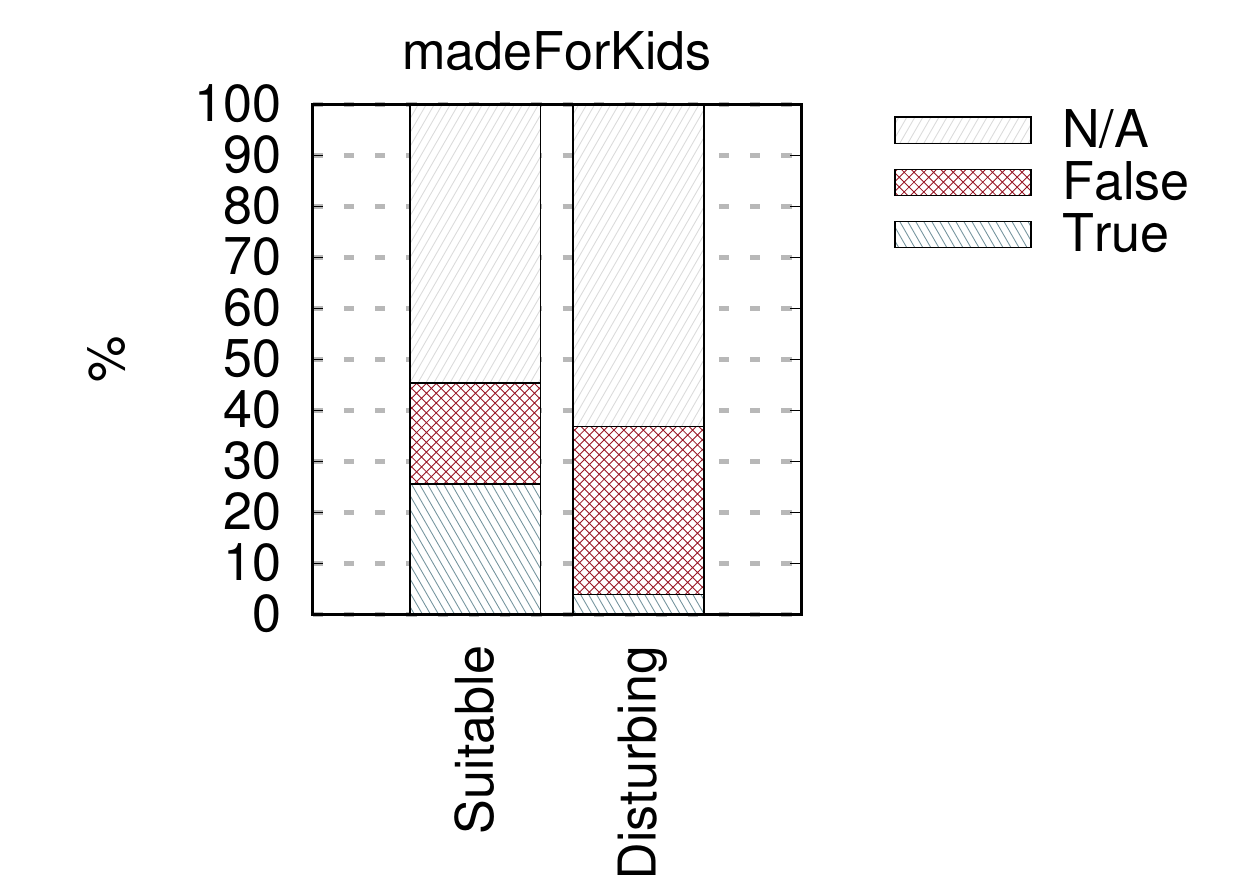}
    \caption{The use of the \emph{madeForKids} label by YouTube channels labeled as suitable or disturbing.}
    \label{fig:madeForKidsChannels}
    \end{minipage}
\end{figure}

Figures~\ref{fig:madeForKidsVideos} and~\ref{fig:madeForKidsChannels} summarize the results of the analysis of the ``madeForKids'' flag, as set by the channel owners.
Given that the videos in $GT$ are targeting kids audience, it comes as no surprise that, as shown in Figure~\ref{fig:madeForKidsVideos}, the majority of videos analyzed are ``madeForKids'', regardless of category, i.e., if they are disturbing or not.
This may be because the creators were aiming to convince the YouTube algorithm that these videos should be recommended to children.
It is encouraging that more suitable videos were marked as ``madeForKids'' than disturbing videos.
Also, out of 390 disturbing videos that were removed, only 1.5\% were set to ``madeForKids''.
Perhaps surprisingly, and according to Figure~\ref{fig:madeForKidsChannels}, most of the channels are not set to ``madeForKids'', even though they hosted such content, possibly because they did not share only such content.
Overall, we find 199 ($\sim$25\%) suitable channels that are exclusively declared as ``madeForKids'', while 3\% of disturbing channels were so.
This may indicate that either the channels posting disturbing videos do not want to draw attention and fast auditing of their videos by YouTube, or their target audience is not kids, and any viewing of their content by kids is accidental.
In either case, we believe there is a significant problem at hand, since kids can reach these videos and channels quite easily, with a few clicks, as shown by past research~\cite{papadamou2020disturbed-youtube-for-kids,papadamou2021characterizing}.

\subsection{Characteristics of YouTube Channels Hosting Videos For Kids}
\label{sec:channel-properties}

\begin{table}[t]
    \caption{Statistics for YouTube channels annotated as suitable or disturbing.}
 \small
    \begin{tabular}{lrrrr}
    \toprule
    \multirow{2}{*}{\bf Features} &
          \multicolumn{2}{c}{\bf Suitable} &
          \multicolumn{2}{c}{\bf Disturbing} \\
    & \bf  Records & \bf Median & \bf Records & \bf Median \\
    \midrule
    videoCount              & 779    & 202     & 559    & 61\\
    viewCount               & 779    & 60M    & 559    & 2488k\\
    subscriptionCount       & 779    & 0   & 559    & 0\\
    subscriberCount         & 700    & 348k      & 524    & 9.7k\\
    \midrule
    descriptionCharCount    & 623    & 287     & 419    & 187\\
    keywordsCount           & 547    & 12      & 312    & 9\\
    topicCount              & 756    & 3   & 524    & 3.0\\
    postCount               & 468    & 2   & 357    & 4\\
    \bottomrule
\end{tabular}
\label{tab:suitable-disturbing-stats}\vspace{-0.3cm}
\end{table}

Next, we analyze the data collected on attributes of each channel, to understand the differences between channels that post only \emph{suitable} videos and those that upload \emph{disturbing} videos.

\begin{figure}[t]
\begin{minipage}[t]{\columnwidth}
    \centering
    \includegraphics[width=0.67\columnwidth]{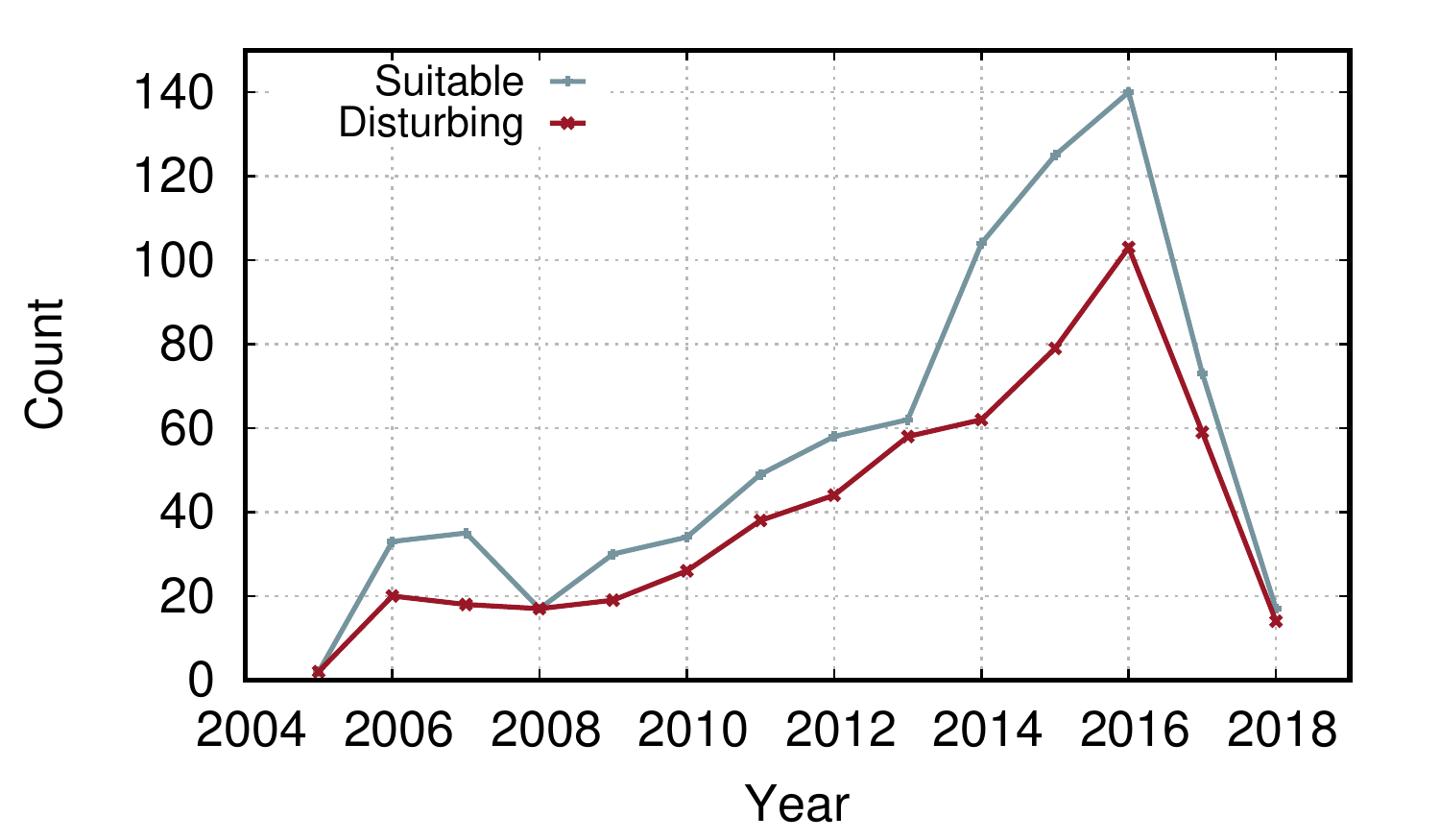}
    \caption{A frequency distribution of the year that YouTube channels were created (channel feature ``publishedAt''), and are labeled as ``suitable'' or ``disturbing''.}
    \label{fig:time}
    \end{minipage}
\begin{minipage}[t]{\columnwidth}
    \centering
    \includegraphics[width=0.67\columnwidth]{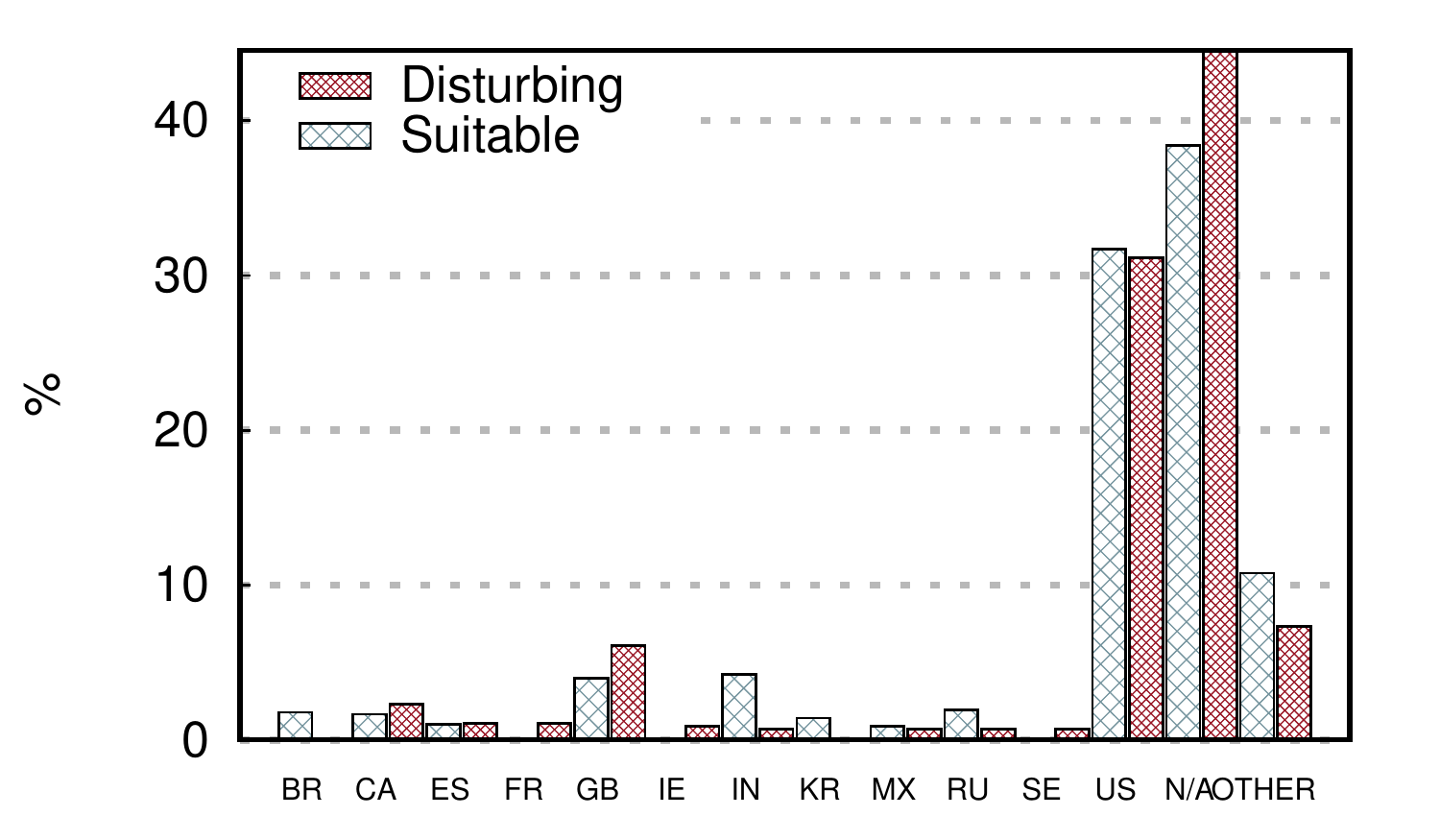}
    \caption{Top 13 countries of channel creation.
    ``N/A'' refers to channels that specifically did not define country.
    ``Other'' refers to channels in countries beyond the top 13 shown here.}
    \label{fig:countries}
    \end{minipage}
\end{figure}

\point{Channel Date Creation, Country and Email}
First, we examine the date (year) channels joined YouTube.
As seen in Figure~\ref{fig:time}, the peak of channel creations for both \emph{disturbing} and \emph{suitable} channels in our dataset is observed in 2016.
After that point, there is a steep decrease in count.
This is due to several measures taken since 2017.
As the term ``Elsagate'' grew popular, Twitter users drew attention on the topic, and in June 2017, a subreddit r/Elsagate~\cite{elsagate-subreddit} was created to discuss and report the phenomenon.
In addition, during the same year, several articles were published about channels featuring inappropriate content and how harmful videos manage to get through the countermeasures of YouTube.
To resolve the controversy, YouTube began to take action by deleting accounts and videos and tightening up its Community policies and guidelines~\cite{elsagate}.

Next, we look into the country of origin which is displayed in the ``Details'', along with ``Email for Business inquires'', in case it exists.
In Figure~\ref{fig:countries}, we plot the top countries that channel owners featured, as well as ``N/A'' for channels that did not display this information.
As perhaps expected, most of the channels originate from United States, with the top 3 popular channels (ranked based on subscribers) being ``Cocomelon'' ($>$100M), ``Kids Diana Show'' and ``Like Nastya'', ranging between 70 and 90M, which are classified as ``suitable'' channels.
It should be noted that an important quantity of \emph{suitable} channels have set their location to India, which is not as frequent in the opposing category (\emph{disturbing}).
Most popular suitable accounts from India include ``ChuChu TV Nursery Rhymes \& Kids Songs'' (46.2M), ``Wow Kidz'' (21.9M), and ``Green Gold TV - Official Channel'' (15.4M).

\begin{figure*}[t]
\begin{minipage}[t]{0.234\textwidth}
    \centering
    \includegraphics[width=\linewidth]{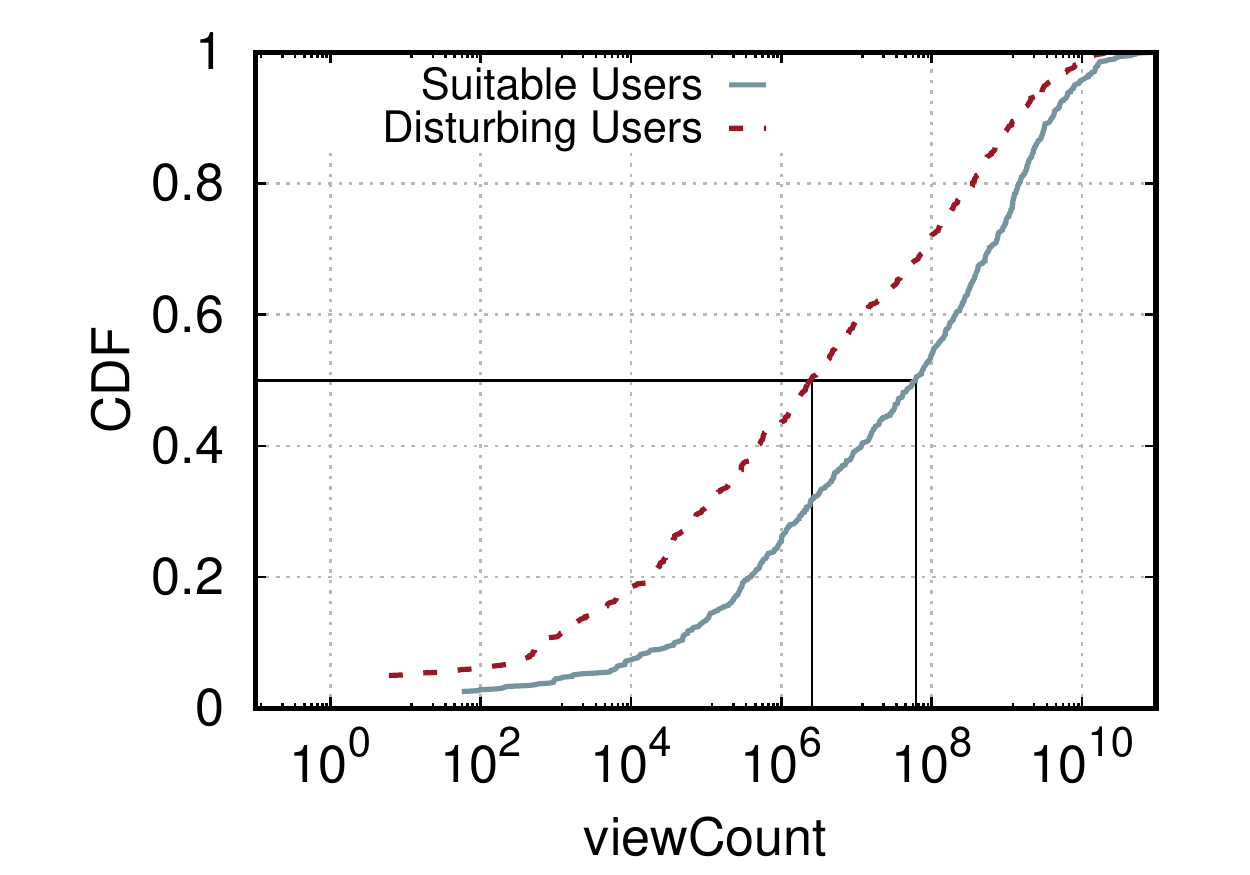}
    \caption{CDF for viewCount (number of total views)  per channel for disturbing or suitable users.}
    \label{fig:viewCount}
\end{minipage}%
\hfill
\begin{minipage}[t]{0.234\textwidth}
    \centering
    \includegraphics[width=\linewidth]{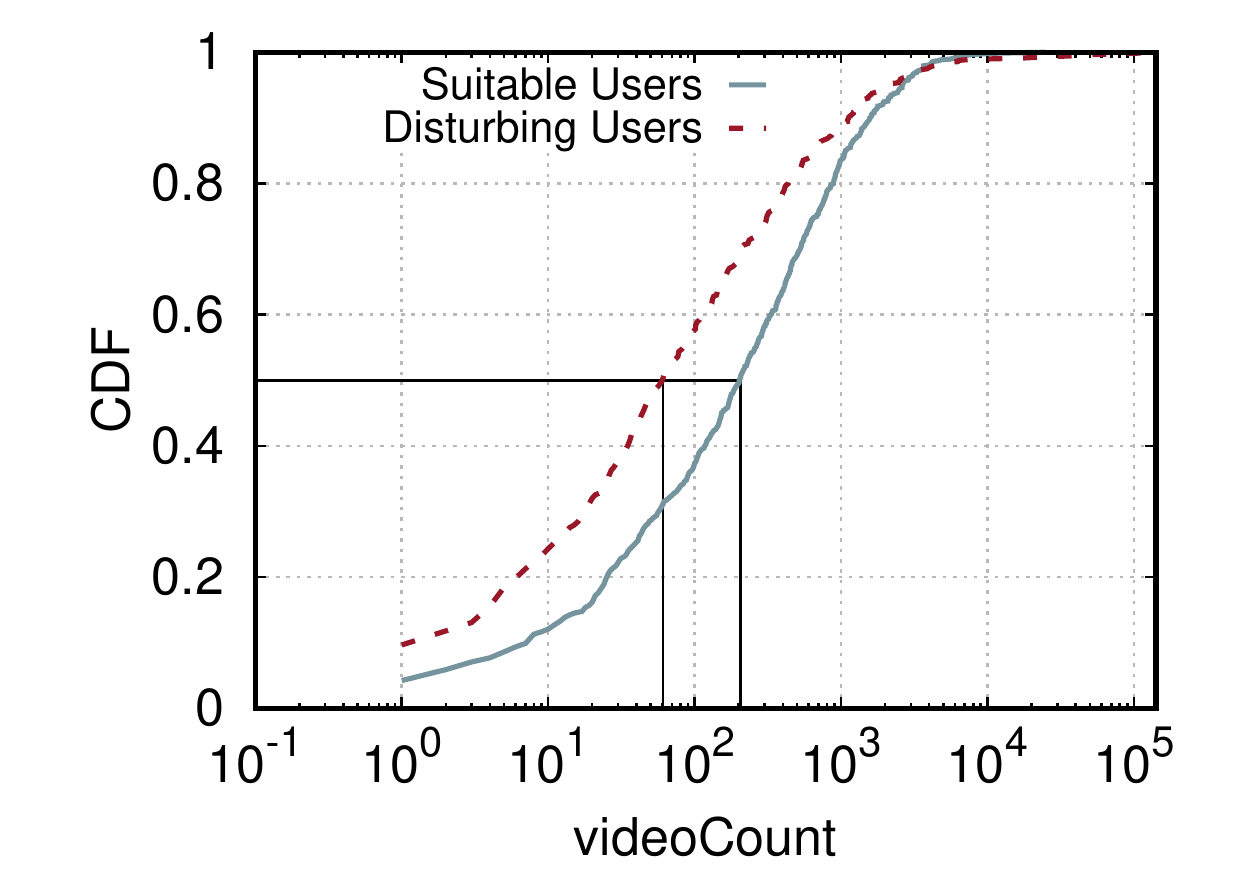}
    \caption{CDF for videoCount (number of current publicly visible videos)  per channel for disturbing or suitable users.}
    \label{fig:videoCount}   
\end{minipage}
\hfill
\begin{minipage}[t]{0.234\textwidth}
    \centering
    \includegraphics[width=\linewidth]{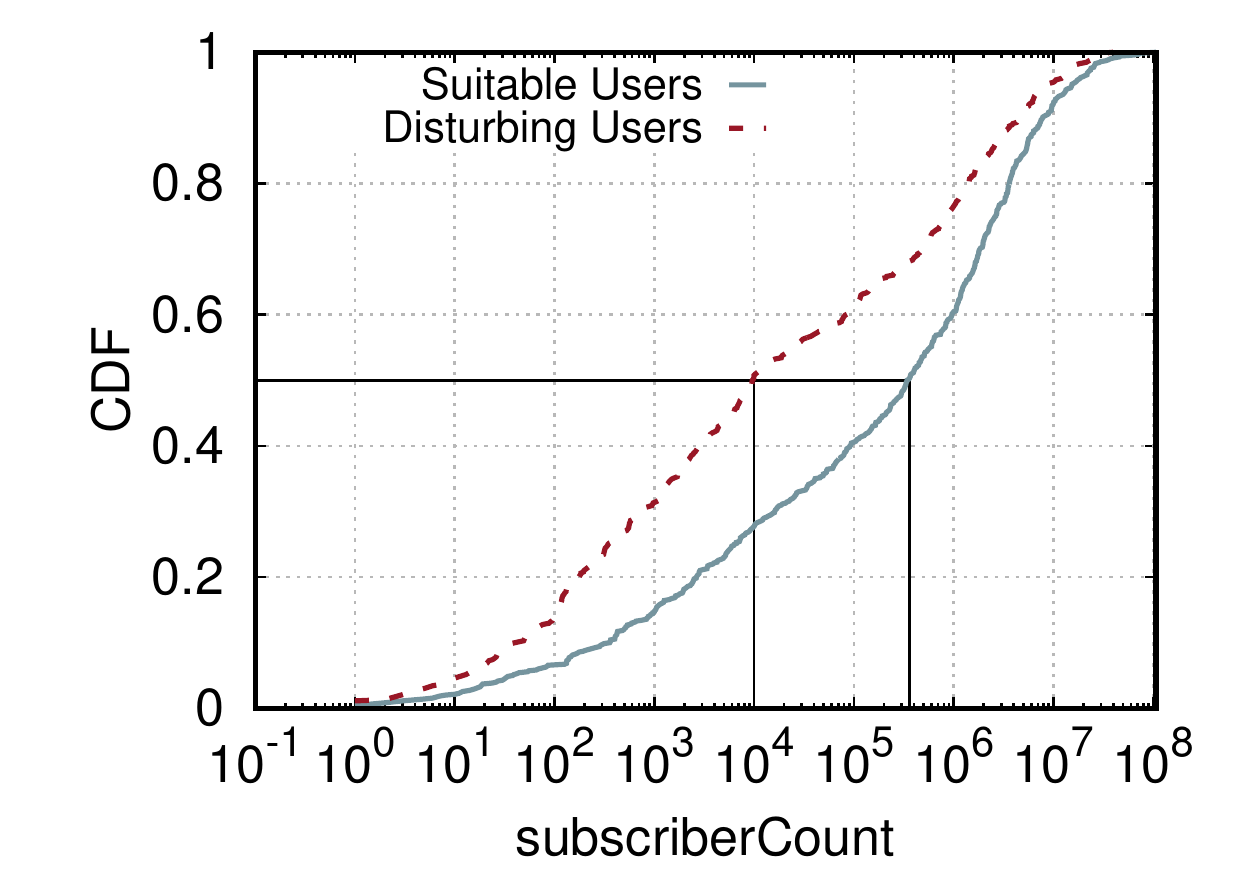}
    \caption{CDF for subscriberCount (can be hidden) per channel for disturbing or suitable users.}
    \label{fig:subscriberCount}
\end{minipage}
\hfill
\begin{minipage}[t]{0.234\textwidth}
    \centering
    \includegraphics[width=\linewidth]{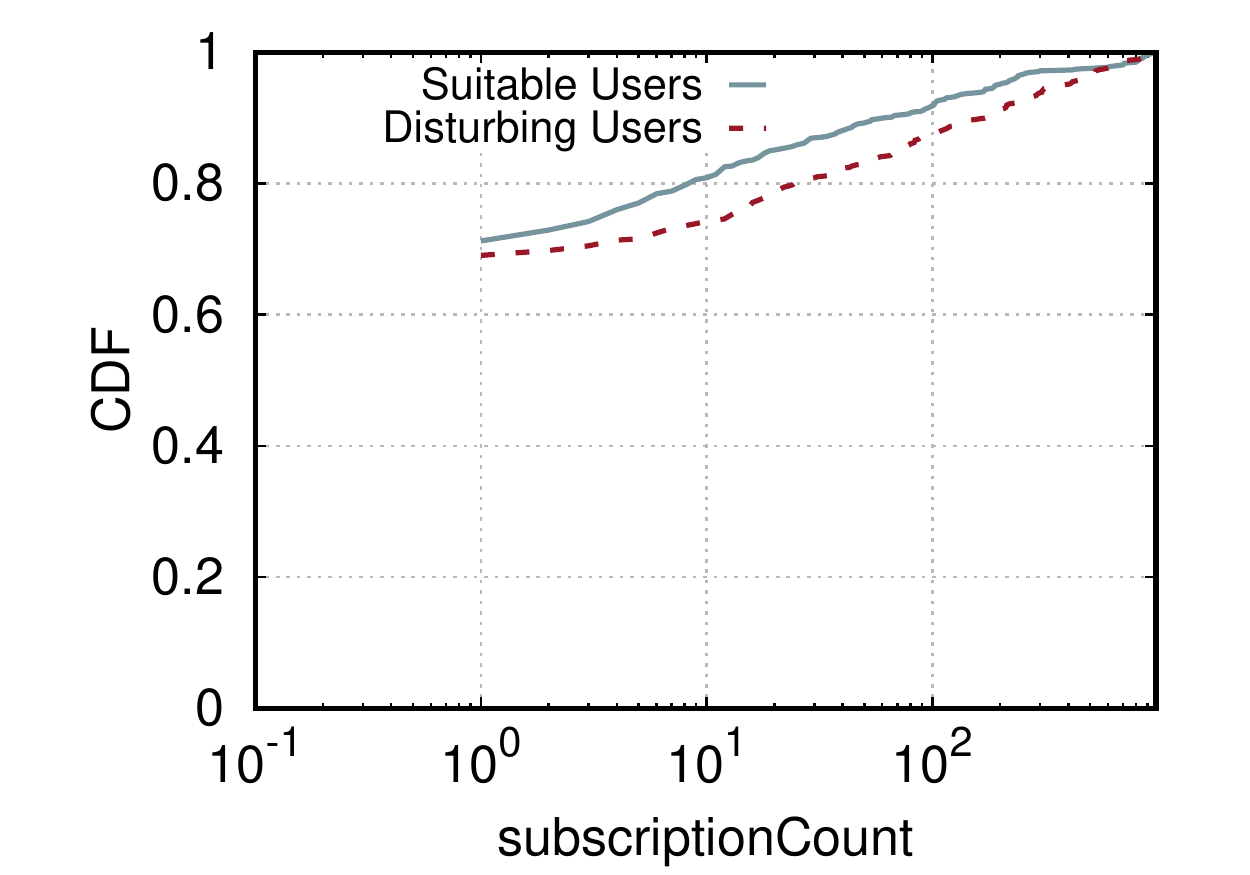}
    \caption{CDF for subscriptionCount (visible subscriptions) per channel for disturbing or suitable users.}
    \label{fig:subscriptionCount}
\end{minipage}
\end{figure*}

\begin{figure*}[t]
    \centering
    \begin{minipage}[t]{0.234\textwidth}
    \centering
    \includegraphics[width=\linewidth]{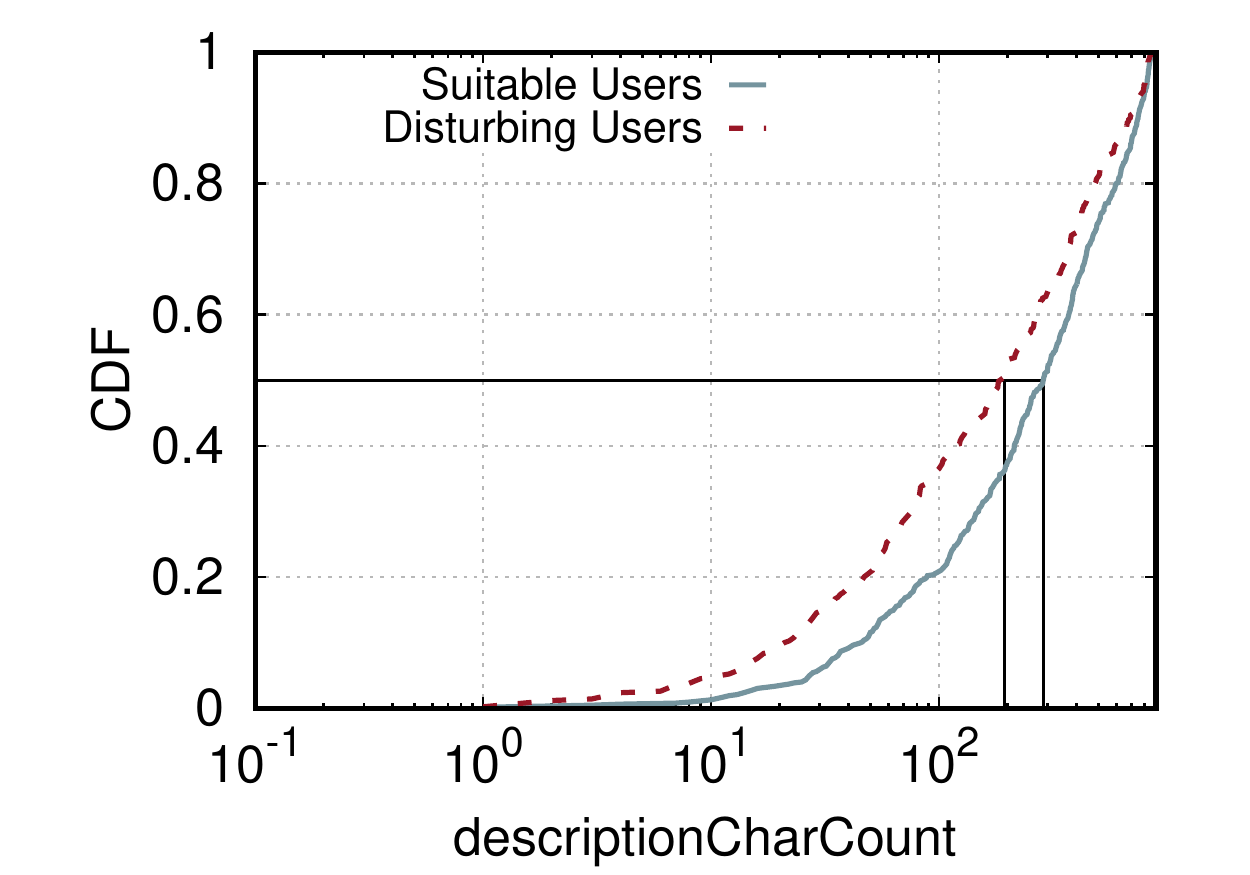}
    \caption{CDF of descriptionCharCount (number of characters in channel description (no spaces)) per channel for disturbing or suitable users.}
    \label{fig:descriptionCharCount}
\end{minipage}%
\hfill
    \begin{minipage}[t]{0.234\textwidth}
    \centering
    \includegraphics[width=\linewidth]{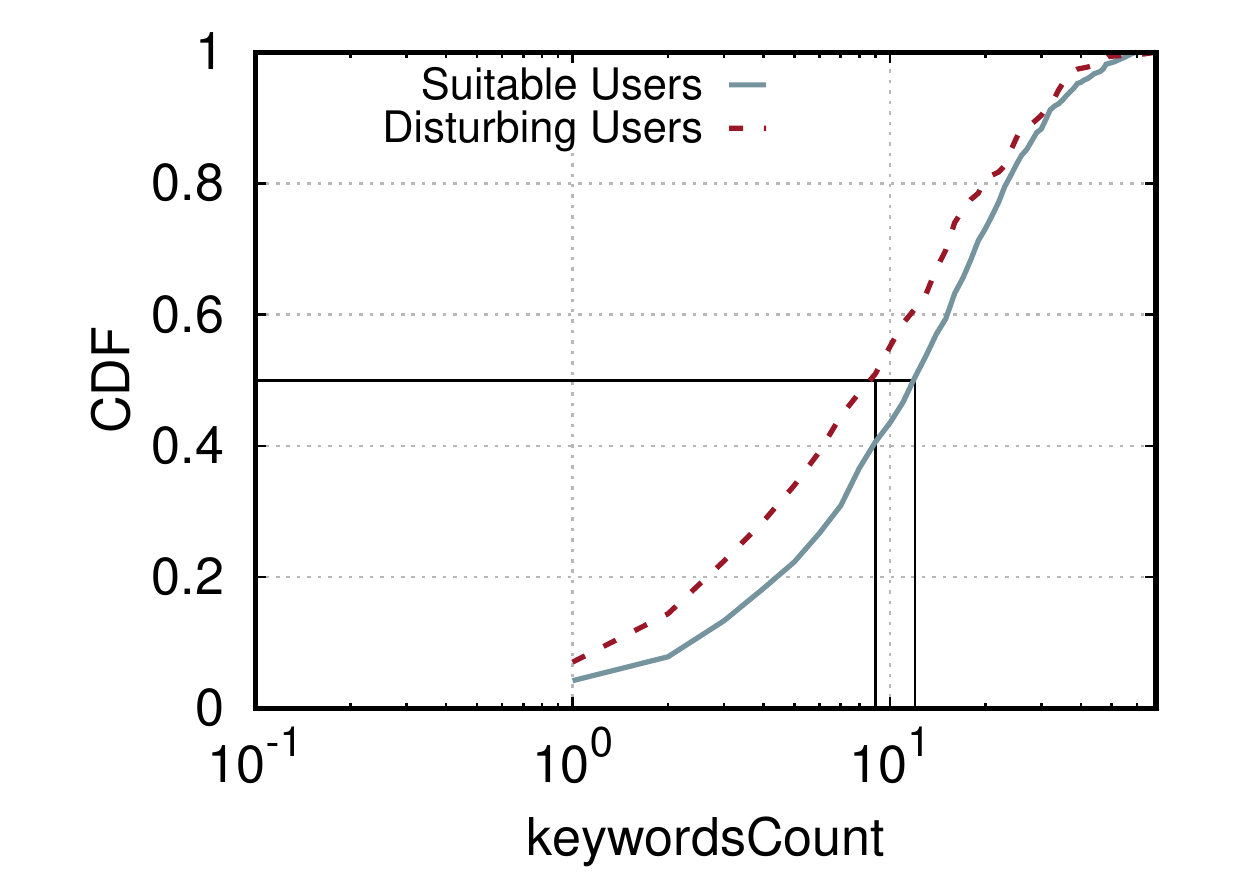}
    \caption{CDF of keywordsCount (number of keywords) per channel for disturbing or suitable users.}
    \label{fig:keywordsCount}
\end{minipage}
\hfill
    \begin{minipage}[t]{0.234\textwidth}
    \centering
    \includegraphics[width=\linewidth]{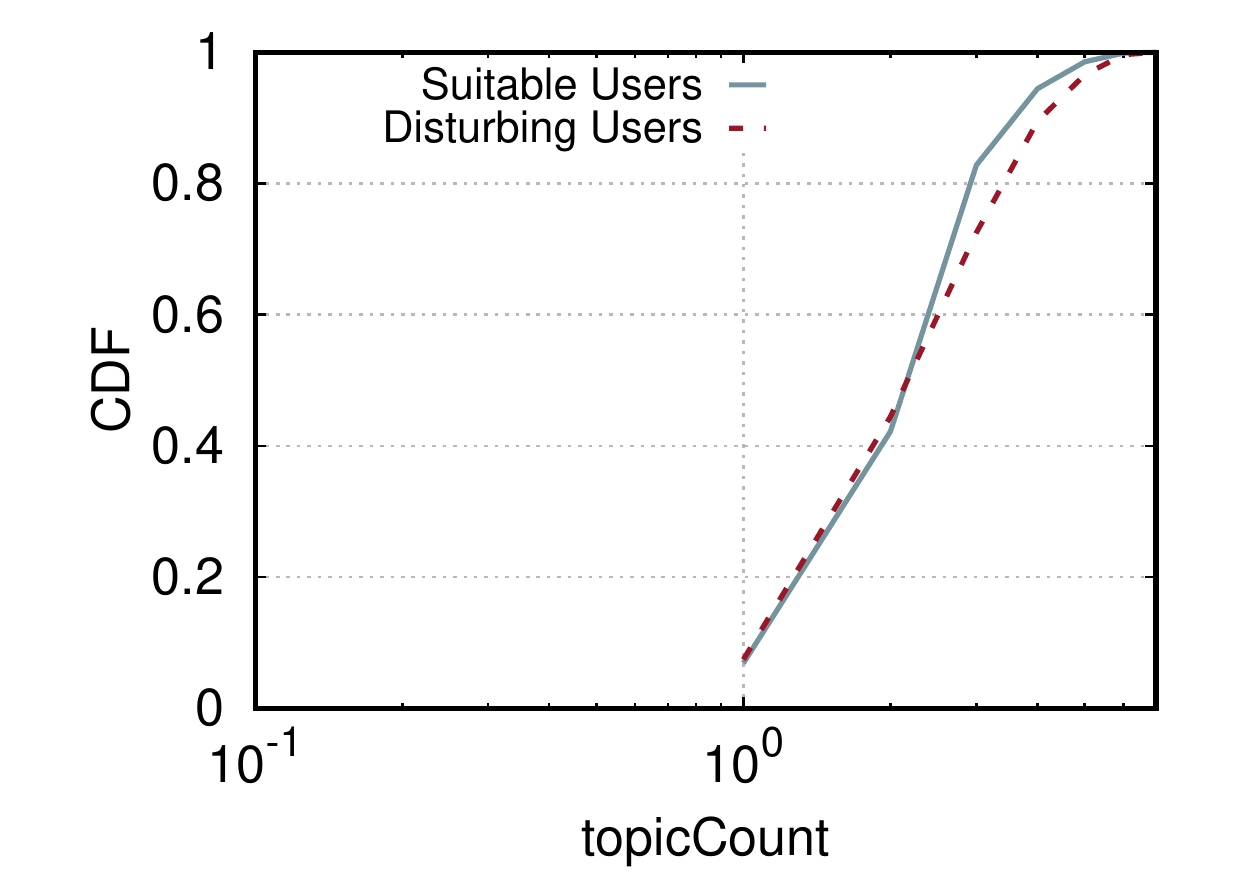}
    \caption{CDF of topicCount (number of topics per channel -- can be hidden)  per channel for disturbing or suitable users.}
    \label{fig:topicCount}
\end{minipage}
\hfill
\begin{minipage}[t]{0.234\textwidth}
    \centering
    \includegraphics[width=\linewidth]{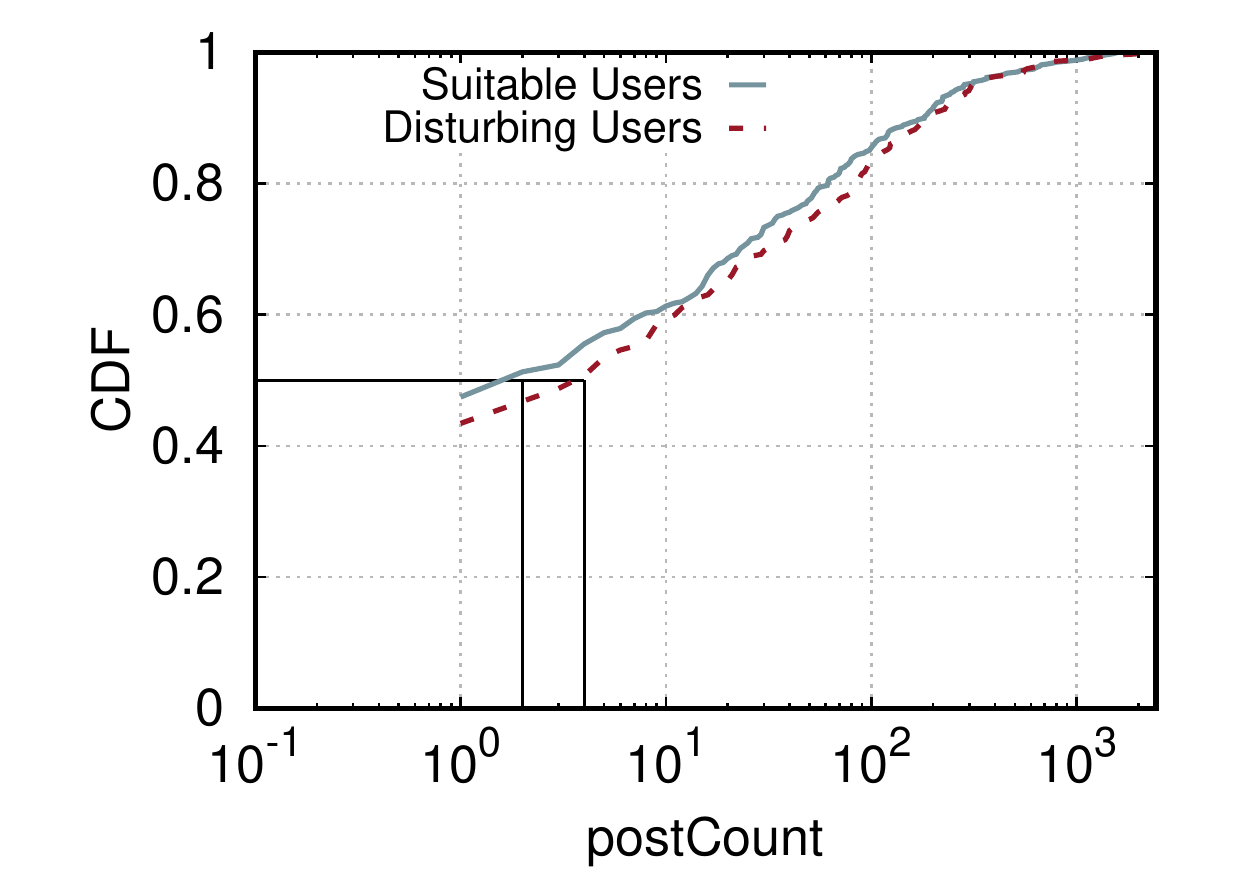}
    \caption{CDF of postCount (number of posts) per channel for disturbing or suitable users.}
    % \caption{ECDF postCount}
    \label{fig:postCount}
\end{minipage}%
\end{figure*}

\point{Channel Statistics and Subscriptions}
Next, we perform non-parametric, Kolmogorov-Smirnov (KS) testing to find out whether or not the distributions of the two types of channels are statistically different.
To begin with, we study the channel statistics, \ie~viewCount, videoCount, subscriberCount and subscriptionCount.
From Figure~\ref{fig:viewCount}, it is evident that suitable channels have more views, on average, than disturbing channels ($\sim$1.7B vs. $\sim$663M).
This is also true for number of videos publicly available on each channel (Figure~\ref{fig:videoCount}), number of subscribers per channel (Figure~\ref{fig:subscriberCount}) and number of subscriptions (Figure~\ref{fig:subscriptionCount}).
It should also be pointed out that the average ratio of views per video is three times higher in channels of suitable than disturbing content (4.2M vs. 1.4M).
Then, as summarized in Table~\ref{tab:suitable-disturbing-stats} for the two type of channels, we look closer into the subscriberCount, which indicates how many people have subscribed to a channel to keep up with its newly updated content and support the creator.
The public subscriberCount is rounded depending on the number of subscribers.
Thus, it is different from the actual subscriber count which is private and exclusively available to the owner of the YouTube channel~\cite{subscribers}.
We collected public subscribersCount for each channel via YouTube Data v3 API.
However, each creator has the option to hide the subscriberCount of their channel.
We observe that $\sim$10\% of \emph{suitable}, but only $\sim$6\% of \emph{disturbing} channels choose to conceal the number of their subscribers.
KS test performed on each of these four features allows us to reject the null hypothesis that the two types of channels originate from the same distribution at statistical level $p-value<0.0328$ or lower (all statistics are summarized in Table~\ref{tab:smirnov}).

\begin{table}[t]
    \caption{Kolmogorov-Smirnov for count-based channel characteristics.}
    \small
    \begin{tabular}{lrr}
        \toprule
        \textbf{Feature}   &\textbf{p-value}   &\textbf{D-statistic}\\ 
        \midrule
        videoCount              & 2.636e-06    & 0.21333\\
        viewCount               & 1.211e-03    & 0.20359\\
        subscriptionCount       & 3.288e-02    & 0.07944\\
        subscriberCount         & 8.882e-15    & 0.23482\\
        \midrule
        descriptionCharCount    & 3.835e-12    & 0.16439\\
        keywordsCount           & 2.729e-13    & 0.13655\\
        topicCount              & 2.867e-03    & 0.10285\\
        postCount               & 6.802e-01    & 0.05049\\
        \bottomrule
    \end{tabular}
    \label{tab:smirnov}\vspace{-0.3cm}
\end{table}

\point{Branding settings, Topic Details and Posts}
Next, we examine the attributes that are related to the content description, \ie~descriptionCharCount, keywordsCount, topicCount, and postCount.
Again, channels with only \emph{suitable} videos seem to have longer descriptions (Figure~\ref{fig:descriptionCharCount}) and more keywords (Figure~\ref{fig:keywordsCount}) used in their configurations.
Interestingly, the distribution of number of topics (Figure~\ref{fig:topicCount}) and number of posts per channel (Figure~\ref{fig:postCount}) seem to be similar for the two types of channels.
As earlier, we performed KS tests and found that we cannot reject the null hypothesis for the postCount feature, and the two types of channels come from the same distribution ($p-value=0.6802$).

\point{Topic Categories and Keywords}
Topic categories and keywords are used to describe and associate a creator's content with specific search results and recommendations.
It is of high importance to set up these features properly in order to reach the desired audience and achieve channel growth.
Both of these features can be collected via YouTube API v3.
In Table~\ref{tab:10keywords-categories} we show the top 10 keywords and top 10 topics used, respectively, for the two types of channels.
It is evident that, apart from the usual children-associated tags which appear to be prevalent on both types of channels, \emph{disturbing} channels use gaming-related keywords and topics more often than \emph{suitable} channels. This is a result of channels uploading MLG~\cite{mlg-memes} content and heavily moded ROBLOX~\cite{roblox} and Minecraft~\cite{minecraft} videos.

\begin{table}[t]
\caption{Ten most used keywords and topicCategories per channel type.}
\centering
\footnotesize
\resizebox{\columnwidth}{!}{%
\begin{tabular}{lll}
\toprule
\textbf{Category}   &\textbf{Keywords (frequency)} &\textbf{topicCategories (frequency)}\\
\midrule
suitable    &kids(70), fun(30), toys(47),       &Entertainment(470), Film(338)\\ 
            &animation(44), children(41),       &Lifestyle\_(sociology)(327), Hobby(221)\\
            &cartoon(34), funny(30),            &Music(185), Television\_program(110)\\
            &cartoons(30), for kids(30)         &Video\_game\_culture(87)\\
            &nursery rhymes(35)                 &Action-adventure\_game(51)\\
            &                                   &Action\_game(50)\\
            &                                   &Role-playing\_video\_game(44)\\
\midrule
disturbing  &funny(47), animation(34),          &Entertainment(343), Film(229)\\ 
            &comedy(26), gaming(18),            &Video\_game\_culture(135), Music(120)\\
            &cartoon(15), kids(15),             &Action-adventure\_game(51)\\
            &cartoons(14), fun(16),             &Action\_game(91)\\
            &minecraft(12), Gaming(11)          &Role-playing\_video\_game(61)\\
            &                                   &Hobby(61), Pop\_music(37)\\
\bottomrule
\end{tabular}%
}
\label{tab:10keywords-categories}\vspace{-0.3cm}
\end{table}

\subsection{Viewers Interaction \& Social Media Presence}
\label{sec:channel-interactions}

Apart from the general features that compose a channel, there are additional capabilities that focus on bridging the connection between a channel and its subscribers. 
Community Tab, which is one of the latest features offered by YouTube, released its beta version in 2016~\cite{community-tab-beta}.
A creator unlocks this feature upon reaching 1000 subscribers, and they can make use of it only if their channel is not set to ``madeForKids''~\cite{madeforkids-coppa}.
From that point on, they are able to create posts and embed playlists, GIFs, images, videos, polls, etc~\cite{community-tab}.
Also, viewers get Community post notifications as they get from video uploads, but only in case their notifications are enabled.

Indeed, a large number of suitable channels do not have the Community Tab feature enabled, as, also pointed out in Section~\ref{sec:made-for-kids}, more than 25\% suitable channels are ``madeForKids''.
Thus, even though they have a higher average number of subscribers than disturbing channels (as was shown in Figure~\ref{fig:subscriberCount}), a significant portion of these channels cannot use the Community Tab feature.
Interestingly, in Figure~\ref{fig:postCount}, disturbing channels exhibit more posts per channel on average than suitable channels.

Channel owners can also display their social media and link their channels to other platforms and websites.
This is shown in the About Tab, which contains general details about a channel. 
More specifically, it includes the channel description, statistics such as date of creation and total views, links and e-mail information.
For each channel, we collected the social media, external URLs and e-mail associated with the account.

\begin{figure}[t]
\begin{minipage}[t]{0.67\columnwidth}
    \centering
    \includegraphics[width=\linewidth]{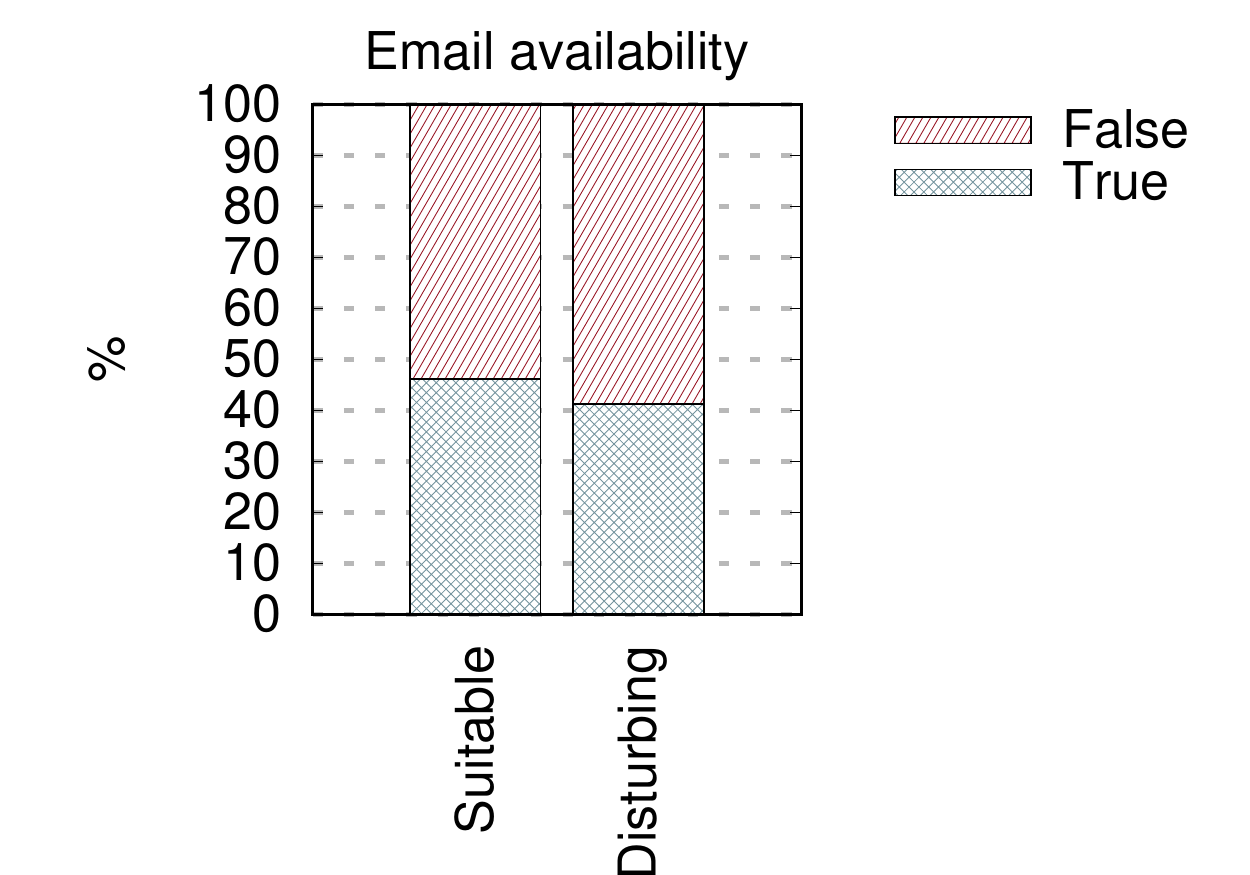} 
    \caption{Use of email for business inquiries in YouTube channels labeled as disturbing or suitable.}
    \label{fig:mail}
    \end{minipage}
    \hfill
    \begin{minipage}[t]{0.67\columnwidth}
    \centering
    \includegraphics[width=\linewidth]{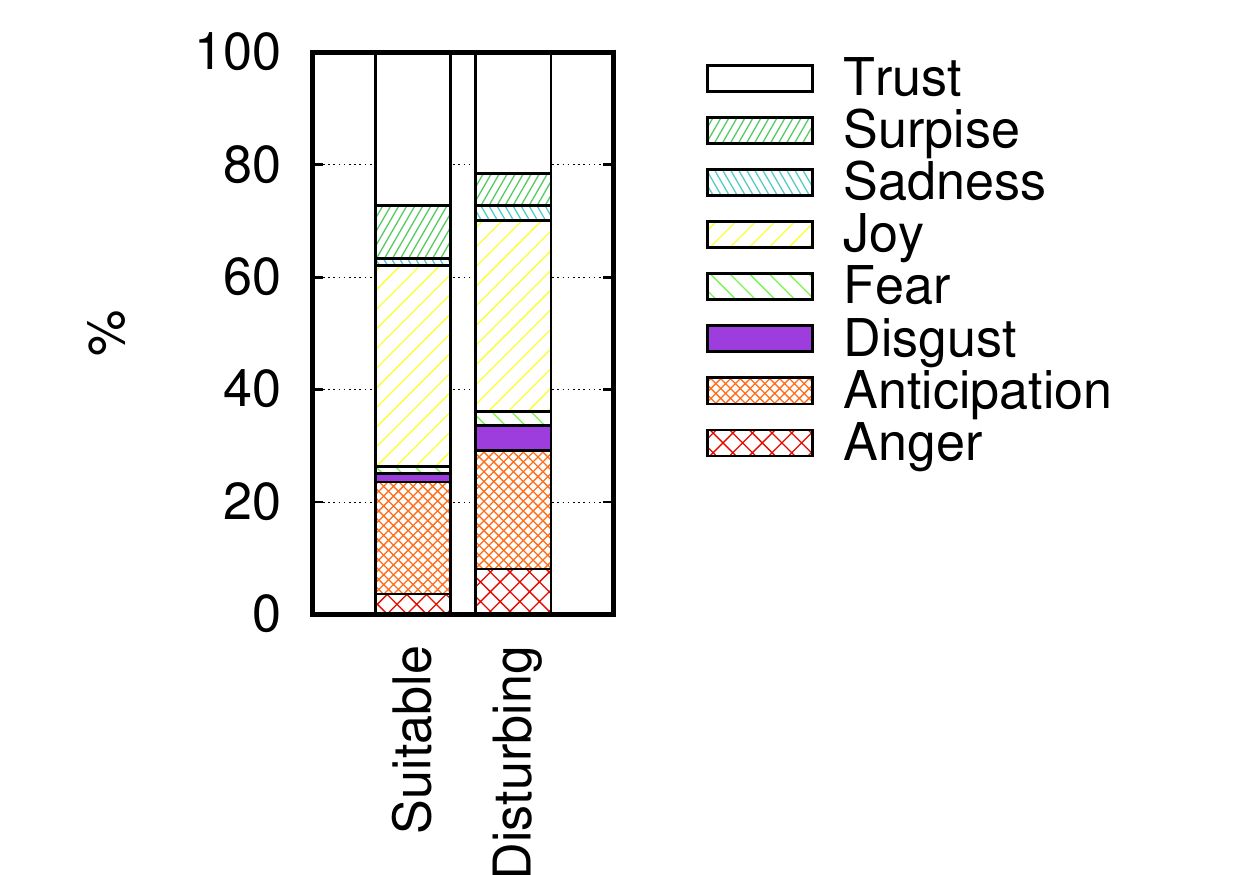} 
    \caption{Emotion analysis on YouTube channel description, for channels labeled as suitable or disturbing.}
    \label{fig:emotionwheel}
\end{minipage}
\end{figure}

The 10 most frequent social media referenced in the About section are shown in Table~\ref{tab:socialmedia}.
As expected, popular networks such as Instagram, Twitter and Facebook are prevalent.
The majority of suitable channels display Facebook in their links, while disturbing channels show a preference for Twitter.
Moreover, by including their contact info, channel owners encourage communication with their audience and are easily accessible for possible collaborations  ~\cite{business-inquiry}.
However, in Figure~\ref{fig:mail}, we see that less that a half of channels for both types provide their email addresses.
Even so, disturbing channels are slightly less likely to add their contact information than suitable channels.

\begin{table}[t]
    \centering
    \caption{Top social media \& websites used or linked in YouTube channels.}
    \small
    \begin{tabular}{lrr}
    \toprule
    \textbf{Platform}   &\textbf{Suitable}  &\textbf{Disturbing}  \\
    \midrule
    facebook    &282 &129 \\
    instagram   &217 &147 \\
    merchandise &16 &25 \\
    twitch      &10 &35 \\
    twitter     &190 &160 \\
    \bottomrule
    \label{tab:socialmedia}
    \end{tabular}\vspace{-0.4cm}    
\end{table}

\begin{figure*}[t]
\centering
\begin{minipage}[b]{0.32\textwidth}
    \centering
    \includegraphics[width=\linewidth]{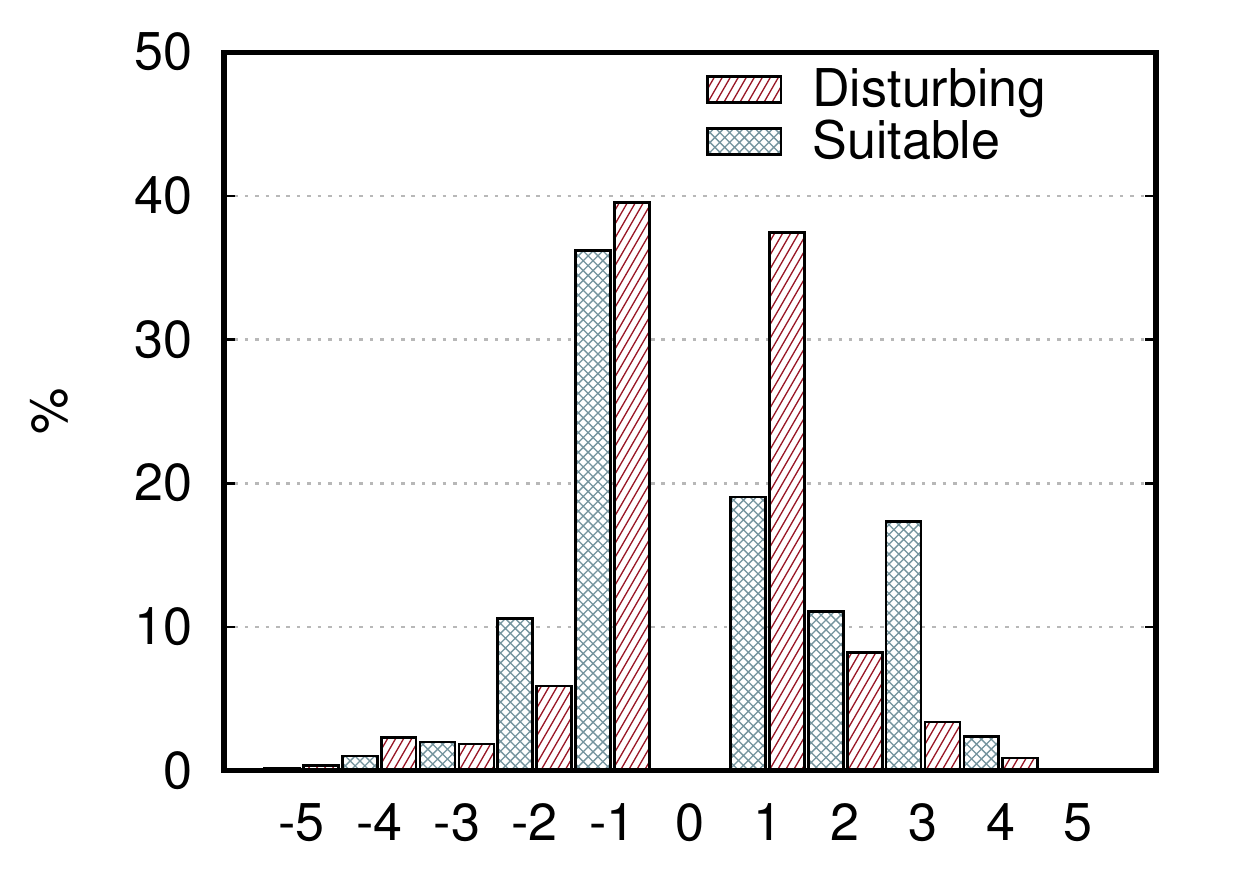}
    \caption{Polarity of description of YouTube channels labeled as suitable or disturbing.}
    \label{fig:descriptionSentistrength}
\end{minipage}%
\hfill
\begin{minipage}[b]{0.32\textwidth}
    \centering
    \includegraphics[width=\linewidth]{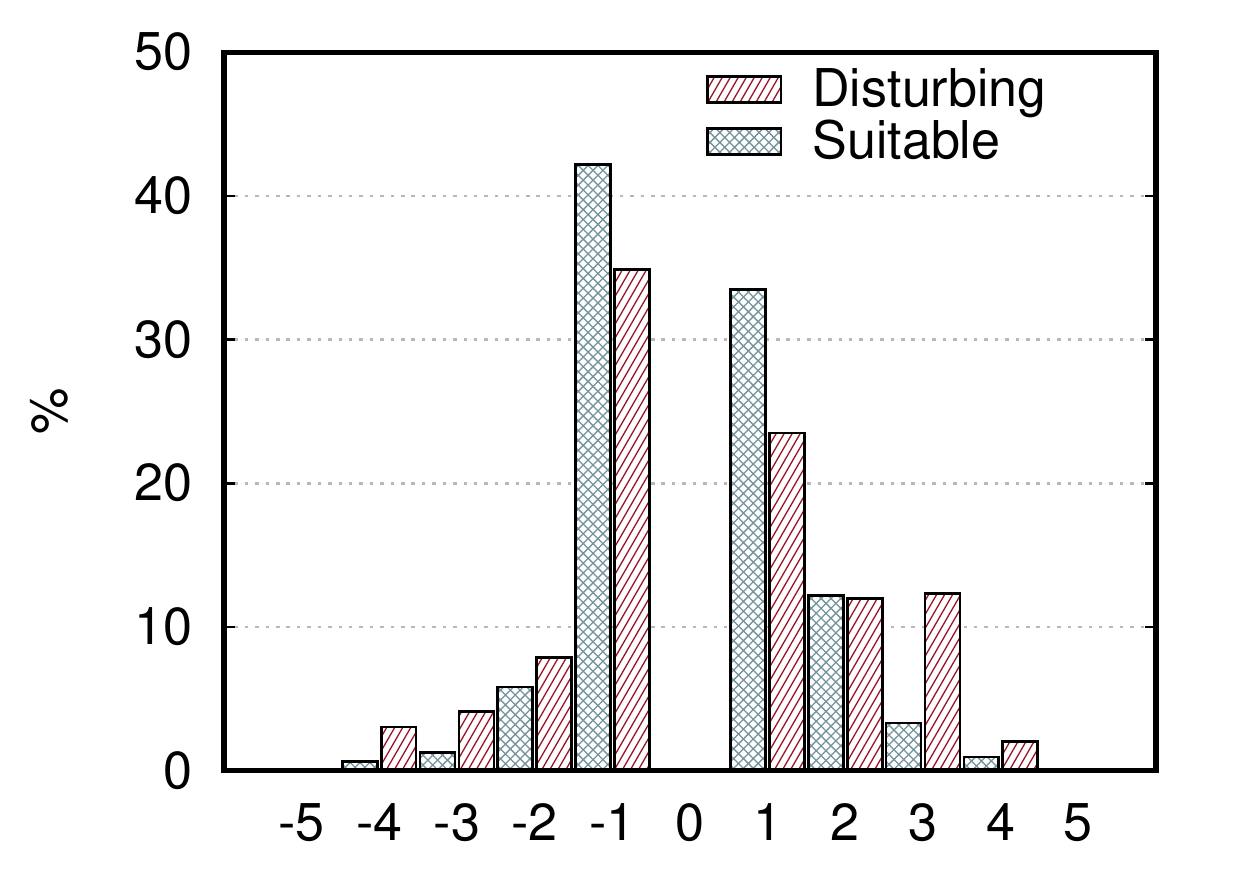}
    \caption{Polarity of keywords of YouTube channels labeled as suitable or disturbing.}
    \label{fig:keywordsSentristrength}
\end{minipage}%
\hfill
\begin{minipage}[b]{0.32\textwidth}
    \centering
    \includegraphics[width=\linewidth]{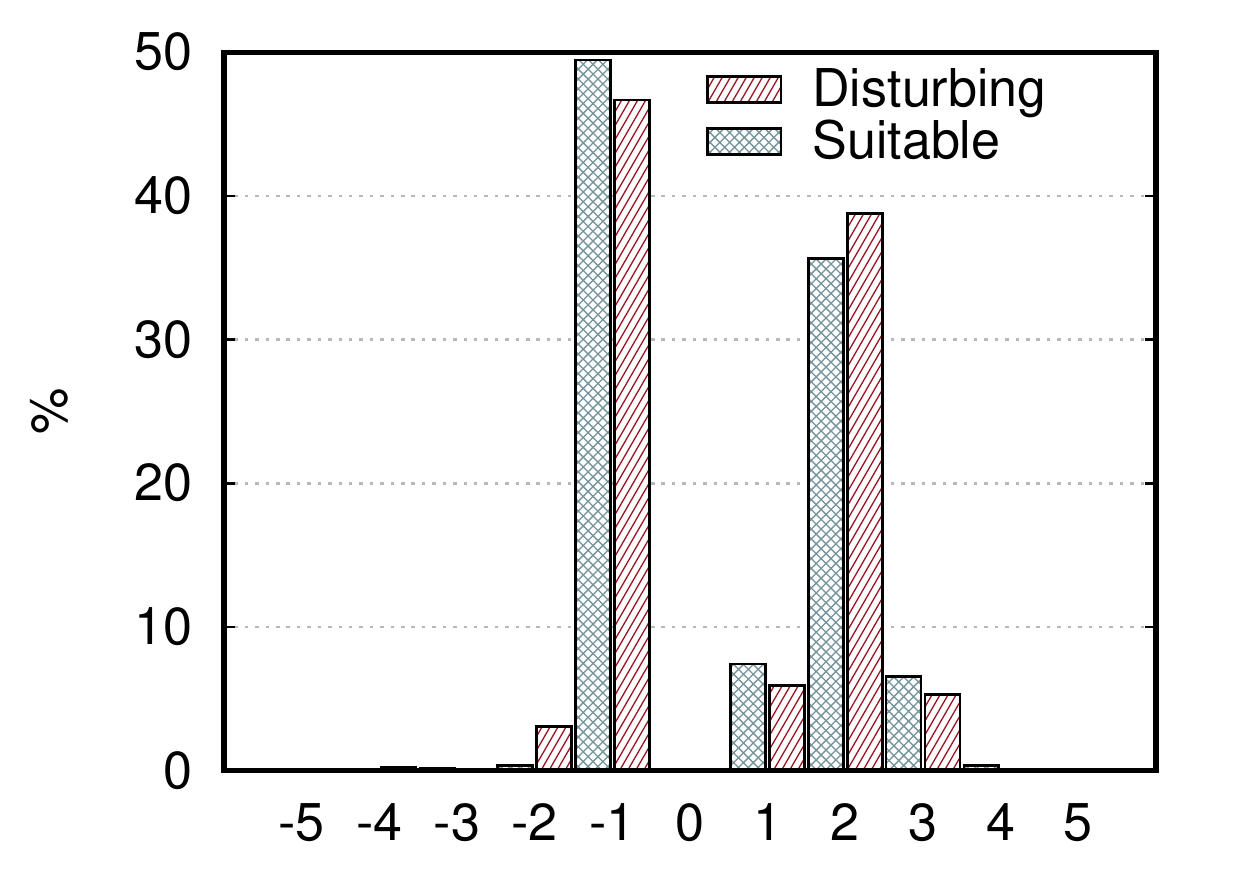}
    \caption{Polarity of posts (mean) of YouTube channels labeled as suitable or disturbing.}
    \label{fig:postsSentristrength}
    \end{minipage}
\end{figure*}

\begin{table}[t]
\caption{Top emoticons used Community Tab posts for YouTube channels labeled as suitable or disturbing.}
\centering
\footnotesize
\begin{tabular}{@{}ccc|ccc@{}}
    \toprule
    \multicolumn{6}{c}{\bf Posts Text} \\
    \midrule
    \multicolumn{3}{c}{\bf Suitable}  & \multicolumn{3}{c}{\bf Disturbing}  \\
    \midrule
    \bf Emoji & \bf Count & \bf Score  & \bf Emoji & \bf Count & \bf Score   \\
    \midrule
    \emojiredheart &19 &0.747        &\emojifacewithtearsofjoy &7 &0.805   \\
    \emojibackhandindexpointingdown &10 &0.557       &\emojiredheart &5 &0.747    \\
    \emojiheartsuit &10 &0.754       &\emojismilingfaceheartshapedeyes &5 &0.765       \\
    \emojismilingfaceheartshapedeyes &10 &	0.765     &\emojiexclamationmark &4 &0.620    \\
    \emojibabychick &10 &0.780      &\emojimalesign &3 &0.580 \\
   \bottomrule
\end{tabular}
\label{tab:emojiposts}
\end{table}

\subsection{Sentiment Analysis}
\label{sec:sentiment}
\point{Basic Emotions}
We present the analysis of sentiment performed on the various data collected per channel that include text, such as the channel keywords and the About and post description.
Beginning with channel description, we conducted analysis on the eight basic emotions as in Robert Plutchik's Wheel of Emotions~\cite{plutchnik-emotions}.
It covers eight prime emotions: Trust, Surprise, Sadness, Joy, Fear, Disgust, Anticipation and Anger.
We use MeaningCloud Emotion Detection Deep Categorization API to extract them.
The results presented in Figure~\ref{fig:emotionwheel} show the percentage of emotion detected in text description.
Negative emotions such as Anger, Disgust, Sadness and Fear are more prevalent in disturbing channels, while positive emotions such as Trust and Surprise are expressed less frequently.
This finding correlates with the nature of disturbing content which is characterized by extreme media content and language.
It is observed that suitable channels' descriptions express more Surprise.
Also, suitable and disturbing channels show similar percentages of Anticipation and Joy.

\point{Polarity}
Then, we look into the positive and negative polarity of the description text, keywords and posts of each channel.
In Figures~\ref{fig:descriptionSentistrength},~\ref{fig:keywordsSentristrength} and ~\ref{fig:postsSentristrength}, we show a breakdown of polarity for each of the previous features.
Regarding the text in their description, both types of channels are using words that convey slightly negative sentiment (-1).
However, disturbing channels’ values are higher than suitable channels, in both negative (-1) and positive (+1) sentiment. 
In fact, for the positive side, the disturbing channels use lightly positive sentiment words (+1) almost twice as much as suitable channels.
Overall, disturbing channels use keywords with higher sentiment than suitable channels, both positive and negative.
This is probably an attempt to evoke attention, emotion and increase possible engagement with the audience.
Similarly, disturbing and suitable channels exhibit a high frequency of lightly negative words (-1) as well as positive words (+2) in their posts.

\point{Emojis} We performed emoji detection in the text of channel description and posts, with the assistance of Python library \emph{emoji}~\cite{python-emoji}.
Tables~\ref{tab:emojiposts} and~\ref{tab:emojidescriptions} show the frequency of emojis and their sentiment score for posts and channel descriptions, respectively, and according to Emoji Sentiment Ranking v1.0~\cite{emoji-sentiment}.
Heart emojis such as \scalebox{0.5}{\emojiheartsuit} and \scalebox{0.5}{\emojiredheart} prevail.
Suitable channels express ownership in their description by using frequently \scalebox{0.5}{\emojicopyright}, \scalebox{0.5}{\emojiregistered} and \scalebox{0.5}{\emojitrademarksign} emojis.
The most frequent emoji in disturbing channels' description is \scalebox{0.5}{\emojibiohazard} (bio-hazard emoji), which even if it does not reflect a specific sentiment score, is associated with negative emotion~\cite{emoji-guide}.

%% file: sections/05_classifier.tex
\section{Disturbing Channel Detection with Machine Learning}
\label{sec:classifier}

\begin{table}[t]
\caption{Top emoticons used in the channel description of YouTube channels labeled as suitable or disturbing.}
\centering
\footnotesize
\begin{tabular}{@{}ccc|ccc@{}}
    \toprule
    \multicolumn{6}{c}{\bf Channel description} \\
    \midrule
    \multicolumn{3}{c}{\bf Suitable}  & \multicolumn{3}{c}{\bf Disturbing}  \\ 
    \midrule
    \bf Emoji & \bf Count & \bf Score  & \bf Emoji & \bf Count & \bf Score \\
    \midrule
    \emojiheartsuit &16 &0.754          &\emojibiohazard &9 &-   \\
    \emojiregistered &15 &0.353         &\emojiheartsuit &8 &0.754\\
    \emojicopyright &15 &0.740          &\emojigreatbritain &4 &-\\
    \emojiredheart &13 &0.747           &\emojiredheart &3 &0.747\\
    \emojitrademarksign &11 &-  &\emojiregistered &3 &0.353\\
  \bottomrule
\end{tabular}
\label{tab:emojidescriptions}
\end{table}

\subsection{Data Preparation \& Performance Metrics}
We use the aforementioned features (also summarized in Table~\ref{tab:features}) to train different classifiers for automatic classification of channels into two classes: 1) likely to post only suitable videos (suitable), 2) likely to post at least one disturbing video (disturbing).
In order to compute the classification task, we performed basic preprocessing of the features available, such as removing features with very little to zero variability, and applying logarithmic transformation on several numeric features for normality purposes.
Table~\ref{tab:features} lists the groups of features used in our classification analysis.
As mentioned earlier, the ``suitable'' channels are 779 and ``disturbing'' channels are 559.
We applied 10-fold cross-validation on the available data, and trained and tested various techniques.
We measured standard ML performance metrics such as True Positive and False Positive Rates, Precision and Recall, F1 score and Area Under the Receiver Operating Curve (AUC).
Where applicable, the scores for these metrics were weighted to take into account individual performance metrics per class.

\subsection{Feature Ranking}

\begin{table}[t]
    \caption{Groups of features used for machine learning classification of channels as suitable or disturbing.}
    \centering
    \footnotesize
    \begin{tabular}{lr}
    \toprule
    \textbf{Group of Attributes}     & \textbf{\# of features}  \\ \midrule
    Channel Details \& Activity (count*)  &              6   \\
    Graph-related metrics (subscriptions, etc.) &                     3   \\
    madeforKids Status (ratios, etc)      &                   4   \\
    Top media linked        &                   11  \\ \midrule
    Top keywords per channel&                   10  \\
    Emotions in Description &                   8   \\
    Top topics on Description&                  11  \\
    Emoji score Posts/Description&              2   \\
    Top emojis in Description&                  10  \\
    Top emojis in Posts     &                   10  \\
    Polarity Posts/Description/Keywords&        6   \\ \bottomrule
    \end{tabular}
    \label{tab:features}\vspace{-0.3cm}
\end{table}

We also performed an analysis of the available attributes, and ranked them based on contribution to the classification task.
In particular, we evaluate the worth of an attribute by measuring the information gain with respect to the class, when each attribute was included or not in the classification task.
This effort was performed with a 10-fold cross validation method, and average scores were computed.
Our analysis shows that the top feature groups are:
\begin{enumerate}[leftmargin=0.5cm]
	\item Polarity (keywords or description)
	\item Channel-statistics metrics such as views, subscriber and video counts, country
	\item Top keywords such as nursery rhymes, children, kids, toys
	\item Top topics such as hobby, game-related, lifestyle
	\item Top emotions on description such as trust, surprise, and anger
	\item Emojis and emoji score in text (description, post text, keywords)
\end{enumerate}
This ranking is in line with the results from the previous section, which highlighted that emotions and polarity of channel description have a different profile in disturbing channels than suitable.
Also, characteristics of the channels such as activity statistics and keywords or topic categories used are significantly different in disturbing than suitable channels.

\subsection{Classifiers Performance}

Table~\ref{tab:classification-results} presents the results achieved with several different classifiers and meta-classifiers.
We find that the typical Random Forest (RF) classifier performs very well across the board, with high True positive and low False positive rates, and higher Precision and Recall than the other classic ML methods.
Only another classifier, meta-classifier (Meta:LogitBoost with RF) which uses a regression scheme as the base learner and can handle multi-class problems, performs somewhat better than simple Random Forest, at the expense of higher computation and memory cost.
Another meta-classifier consisting of 4 others (Random Forest, Logistic Regression, Naive Bayes and Bagging RF) and applying average probabilistic voting among them performs similarly.

Regarding the neural network classifier, we tried different architectures, including dense layers for normalization, dropout, etc.
However, due to the small number of examples available in our dataset (1338 samples), these more complex classifiers did not fare better than the simple architecture reported in the results.

We also attempted to build a RF classifier that uses only the features available at the moment the channel is generated.
That is, we dropped features that stem from user and channel activity such as counts (view, video, subscriptions, etc.), posts and their emotion analysis, etc.
The results shown in the last row of Table~\ref{tab:classification-results} demonstrate that it is in fact possible to predict with good performance which channels are likely to post some disturbing content targeting kids, before they have posted anything in their channel, or had any views or subscribers, etc.

\begin{table*}[t]
    \centering
    \caption{Performance metrics from ML binary classification of channels. 
    0: likely to post suitable only content;
    1: likely to post at least one disturbing video.}
    \begin{tabularx}{\textwidth}{lXXXXXX}
    \toprule
    \bf Method     & \bf   TPRate &  \bf  FPRate &  \bf  Precision &  \bf Recall  & \bf   F1      &  \bf  AUC     \\ \midrule
    Random Forest (RF)      &   0.791   &   0.225   &   0.790   &   0.791   &   0.790   &   0.873   \\
    Logistic Regression (LR)&   0.753   &   0.256   &   0.755   &   0.753   &   0.754   &   0.820   \\
    Naive Bayes (NB)        &   0.716   &   0.321   &   0.713   &   0.716   &   0.712   &   0.786 \\
    Neural Net (38x128x2)   &   0.761   &   0.246   &   0.763   &   0.761   &   0.762   &   0.818 \\ \midrule
     Meta: LogitBoost(RF)   &   0.796   &   0.218   &   0.796   &   0.796   &   0.796   &   0.873      \\
    Meta: AvgProb(RF,LR,NB,BRF) &   0.782   &   0.237   &   0.781   &   0.782   &   0.781   &   0.864   \\ \midrule
    RF with only channel gen. features  &   0.781   &   0.222   &   0.784   &   0.781   &   0.782   &   0.869    \\ \bottomrule
    \end{tabularx}
    \label{tab:classification-results}\vspace{-0.3cm}
\end{table*}

%% file: sections/07_related.tex
\section{Related Work}
\label{sec:rel-work}

Previous studies have been conducted regarding disturbing content targeting children in video and streaming platforms.
Ishikawa et al.~\cite{ishikawa2019elsagate-phenomenon} combined raw frames and MPEG motion vectors as a ground dataset to build a classifier detecting Elsagate-related videos.
They propose various machine learning models and compare their performances, as well as ways to reach into a mobile compatible solution with 92.6\% accuracy.
They also mention the ambiguity of ``Elsagate'' definition, and the danger of false positives of this type of content.
Alghowinem~\cite{alghowinem2019youtube-kids-multimodal-analysis} used slices of videos accompanied with audio analysis and speech recognition to provide a real-time mechanism for monitoring content on YouTube and detect inappropriate content for kids.

Next study of relevance is KidsTube by Kaushal et al.~\cite{kaushal2016Kidstube}.
Initially, the authors studied three major feature layers: video, user and comment data.
Then, they built a classifier on these data, as well as a version that is based on a Convolutional Neural Network that uses video frames.
The popularity and network of content uploaders was examined through user statistics such as subscriptions, views, etc.
In fact, they found that likes, subscriptions and playlists can form a network of unsafe promoters and video uploaders.

Another user-centered study is by Benevenuto et al.~\cite{benevenuto2012spammers-content-promoters} which comments on content pollution in video sharing platforms and provides a classification approach at separating spammers and promoters from appropriate users.
Furthermore, Reddy et al.~\cite{reddy2021kid-friendly-youtube-access-model} suggested an age detection process for underage YouTube users, supported by performing sentiment analysis on comments.
In this way, the authors offer a real time protection mechanism that can be embedded in the current YouTube platform.
Continuing with Alshamrani et al.~\cite{alshamrani2020inappropriate-comments}~\cite{alshamrani2021inappropriate-comments}, they perform analysis of exposure of YouTube users to comments, and construct a classifier to detect inappropriate comments in children-oriented videos.
They find that 11\% of comments posted in such videos are toxic.

Lastly, Papadamou et al.~\cite{papadamou2020disturbed-youtube-for-kids}, collect videos targeting children using various seed keywords from animation movies and popular cartoons.
They analyze various types of features available or constructed per YouTube video, and based on these features, the authors build a classifier with 84.3\% accuracy which detects inappropriate videos that target children.
They also underline the dangers of leaving toddlers to watch YouTube videos unattended, and examine the likelihood of a child browsing the platform and coming across a disturbing video by chance.
Our ground truth dataset originates from this study, from which we use the labels provided per suitable or disturbing video.

\point{Comparison}
Our present study goes beyond the aforementioned past works in the following ways:
\begin{itemize}[leftmargin=0.5cm]
    \item
    We shift the problem of \emph{disturbing} videos into the topic of potentially disturbing users creating this type of content.
    In fact, we are the first to check the status (i.e., if they are available or not) of the videos and channels after an interval of two years, and examine the reasons why they may have been removed by YouTube and in what extent.
    \item We are the first to examine the newly introduced ``madeForKids'' flag for both videos and channels, and how its value associates with the type of channel (suitable or disturbing).
    \item We extract and analyze Community Tab posts and perform sentiment and polarity analysis on channel description and post texts.
    \item Furthermore, we use channel public features (e.g., activity and channel related details, posts, keywords, etc.), as well as features not available from the API (e.g., linked media, top emojis topics, polarity, emotions, etc.), to construct a machine learning classifier which detects with good performance channels likely to share disturbing content.
\end{itemize}

%% file: sections/08_conclusion.tex
\section{Discussion \& Conclusion}
\label{sec:conclusion}

The present study focused on an investigation of YouTube channels with respect to the type of videos they share and if these are classified as \emph{disturbing} or \emph{suitable} for kids.

\point{Findings}
\begin{itemize}[leftmargin=0.5cm]
    \item We looked into whether older videos and accounts have been banned by YouTube for violating its policies on content publishing, and examine the reasons why the channels were removed. Alarmingly, we find that the majority of \emph{disturbing} videos (60\%) from a past study (2019), along with their uploaders (channels) (71\%) are still available in mid 2021, during the time interval that our data collection was performed.
    \item We studied the newly added flag from YouTube called ``madeForKids'' to understand how channels and videos marked as disturbing may be correlated to it. We discovered that 1/4th of channels with suitable content are set to ``madeForKids'', but only 3\% of disturbing channels are set as such, which may stem from efforts to avoid attention from YouTube.
\end{itemize}
Furthermore, we studied 27 publicly available features and examined how they are linked to the type of YouTube channel (i.e., if it was found to solely share suitable videos for kids, or disturbing as well) and made several observations that differentiate channels hosting disturbing from suitable videos for kids.
A list of the most important findings on these features are presented below:
\begin{itemize}[leftmargin=0.5cm]
    \item A large number of channels were created in 2016. After that point, less disturbing channels were created, as ``Elsagate'' started to gain attention in 2017 leading to shutdown of disturbing channels from YouTube.
    \item Suitable channels have higher number of views and subscribers than channels with disturbing videos.
    \item Suitable channels tend to use more keywords and have longer descriptions than disturbing channels.
    \item Disturbing channels use gaming-related keywords and topics more often than the suitable channels.
    \item The majority of suitable channels add Facebook in their links; disturbing channels prefer Twitter.
    \item The majority of channels do not provide their email address. However, disturbing channels are slightly less likely to add their contact information.
    \item Negative emotions such as Anger, Disgust and Sadness are more prevalent in disturbing channels than suitable channels.
    \item Disturbing channels use keywords with higher sentiment, negative or positive, in comparison to suitable channels.
\end{itemize}

\point{Automatic ML Classifier}
Finally, based on these studied features, we constructed machine learning (ML) classifiers which detect with adequate performance (up to $AUC$=$0.873$) channels likely to share disturbing content.
In fact, we show how this classification is possible to be performed even at the time a channel is created, by using only features available at that moment and disregarding their activity history or posting features, with up to $AUC=0.869$.
For reproducibility purposes, we make all our data and code available.

\point{Impact}
We believe our analysis of the ``madeForkids'' flag, the characteristics of the disturbing accounts and the ML-based classifier can be combined with other automated tools readily available by academia and YouTube, to fight against inappropriate content exposure and especially when it is targeting kids.
In particular, YouTube could use the results of this study with respect to features differentiating disturbing and suitable accounts, and our suggestion of an ML-based classifier, to create a multi-step process for flagging channels sharing inappropriate content.
This process can follow these steps:
\begin{itemize}
    \item[Step 1:] Extract detailed features per channel, as explained here.
    \item[Step 2:] Train ML method based on these features to detect accounts posting potentially disturbing videos for kids.
    \item[Step 3:] Extract detailed features per video posted in such accounts, following methodology of~\cite{papadamou2020disturbed-youtube-for-kids}.
    \item[Step 4:] Train ML method based on these features, and use it to detect potentially disturbing videos.
    \item[Step 5:] Rank said accounts from Step 2 based on appropriate metric of disturbing content severity such as: the probability of said accounts being disturbing (based on the ML classifier of Step 2), the probability of said videos being disturbing (based on the ML classifier of Step 4), the number of disturbing accounts posted by said account, etc.
    \item[Step 6:] Human moderators can then look into the top ranked disturbing accounts for potential violation of Terms and Conditions and Community Guidelines of YouTube, and consider applying the 3-strike policy.
\end{itemize}
This process could be used as a safety net when the YouTube for Kids application is not available in the country of residence of the children using YouTube.
\point{Limitations}
Last but not least, we shall not forget to mention the limitations of this research. The dataset size is limited as it strictly consists of channels that have uploaded videos from the previous study. There is a selection bias in the sense that the dataset does not cover the whole YouTube platform, but it emerges from child-related content. 
In addition, from our findings, it is apparent that there is a discrepancy between what YouTube considers inappropriate and worth striking and what humans think of as disturbing. For example, many ``disturbing'' annotated videos may fall into the category of dark or adult humour which does not necessarily mean that they should be punished by the platform moderators. Consequently, it is difficult to decide whether ``disturbing'' videos should be removed or there should be better monitoring or categorization of videos to multiple age levels.

Overall, with our present study, we hope to raise awareness about this problem, and encourage YouTube and other similar video sharing platforms to take appropriate measures for protecting children from abusive, disturbing, and generally inappropriate content.

%% file: sections/appendix.tex
\section{Ethical Considerations}
\label{sec:ethics}

The execution of this work has followed the principles and guidelines of how to perform ethical information research and the use of shared measurement data~\cite{dittrich2012menloreport,rivers2014ethicalresearchstandards}.
In particular, this study paid attention to the following dimensions.

We keep our crawling to a minimum to ensure that we do not slow down or deteriorate the performance of the YouTube service in any way.
Whenever possible, we used the recommended YouTube API v3.
When the data to be crawled were not available by the API, we crawled the channel page directly.
We do not interact with any component in each visited page.
In addition to this, our crawler has been implemented to wait for both the page to fully load and an extra period of time before visiting another page.
Also, we do not share any data collected by our crawler with any other entity.